\documentclass[aps,prb,showpacs,twocolumn]{revtex4-1}
\usepackage{amsmath,amssymb,dcolumn,epsfig,graphicx,longtable}
\begin{document}
\title{Rungs 1 to 4 of DFT Jacob's ladder: extensive test on the lattice constant,
bulk modulus, and cohesive energy of solids}
\author{Fabien Tran}
\author{Julia Stelzl}
\author{Peter Blaha}
\affiliation{Institute of Materials Chemistry, Vienna University of Technology,
Getreidemarkt 9/165-TC, A-1060 Vienna, Austria}

\begin{abstract}

A large panel of old and recently proposed exchange-correlation functionals
belonging to rungs 1 to 4 of Jacob's ladder of density functional theory are
tested (with and without a dispersion correction term) for the calculation of
the lattice constant, bulk modulus, and cohesive energy of solids.
Particular attention will be paid to the functionals
MGGA\_MS2 [J. Sun \textit{et al.}, J. Chem. Phys. \textbf{138}, 044113 (2013)],
mBEEF [J. Wellendorff \textit{et al.}, J. Chem. Phys. \textbf{140}, 144107 (2014)],
and SCAN [J. Sun \textit{et al.}, Phys. Rev. Lett. \textbf{115}, 036402 (2015)]
that are approximations of the meta-generalized gradient type and were developed
with the goal to be universally good.
Another goal is also to determine for which semilocal functionals and groups
of solids it is beneficial (or not necessary) to use the Hartree-Fock exchange
or a dispersion correction term.

\end{abstract}

\maketitle

\section{\label{introduction}Introduction}

Starting in the mid 1980's,\cite{BeckeJCP86,PerdewPRB86} there has been a
constantly growing interest in the development of exchange-correlation (xc)
functionals $E_{\text{xc}}=E_{\text{x}}+E_{\text{c}}$ in
the Kohn-Sham (KS) density functional theory (DFT),
\cite{HohenbergPR64,KohnPR65} and the number of functionals that have been
proposed so far is rather huge (see, e.g., Refs.
\onlinecite{CohenCR12,BurkeJCP12,BeckeJCP14,JonesRMP15} for recent reviews).
This is understandable since KS-DFT is the most used method for the
theoretical modeling of solids, surfaces, and molecules at the quantum level,
and the accuracy of a KS-DFT calculation depends to a large extent on the chosen
approximation for $E_{\text{xc}}$. Over the years, the degree of sophistication
of the functionals (and their accuracy) has increased and most of the functionals
belong to one of the rungs of Jacob's ladder.\cite{PerdewAIP01,PerdewJCP05}
On the first three rungs there are the so-called semilocal (SL) approximations which
consist of a single integral for $E_{\text{xc}}$,
\begin{equation}
E_{\text{xc}}^{\text{SL}} = \int\epsilon_{\text{xc}}(\mathbf{r})d^{3}r,
\label{ExcSL}
\end{equation}
and where $\epsilon_{\text{xc}}$, the exchange-correlation energy density per
volume, is a function of (a) the electron density
$\rho=\sum_{i=1}^{N}\left\vert\psi_{i}\right\vert^{2}$ in the local density
approximation (LDA, first rung), (b) $\rho$ and its first derivative $\nabla\rho$
in the generalized gradient approximation (GGA, second rung), and (c) $\rho$,
$\nabla\rho$, and the kinetic-energy (KE) density
$t=\left(1/2\right)\sum_{i=1}^{N}
\nabla\psi_{i}^{*}\cdot\nabla\psi_{i}$
and/or $\nabla^{2}\rho$ in the meta-GGA approximation (MGGA, third rung).
On the fourth rung there are the functionals which make use of the
(short-range, SR) Hartree-Fock (HF) exchange, like the hybrid functionals
\cite{BeckeJCP93}
\begin{equation}
E_{\text{xc}}^{\text{hybrid}} = E_{\text{xc}}^{\text{SL}} +
\alpha_{\text{x}}\left(E_{\text{x}}^{\text{(SR-)HF}} -
E_{\text{x}}^{\text{(SR-)SL}}\right),
\label{Exchybrid}
\end{equation}
where $\alpha_{\text{x}}$ ($\in [0, 1]$) is the fraction of HF exchange energy
$E_{\text{x}}^{\text{(SR-)HF}}$ which is a double integral:
\begin{eqnarray}
E_{\text{x}}^{\text{(SR-)HF}} & = &
-\frac{1}{2}\sum_{i=1}^{N}\sum_{j=1}^{N}\delta_{\sigma_{i}\sigma_{j}}
\int\int
\psi_{i}^{*}(\mathbf{r})\psi_{j}(\mathbf{r}) \nonumber\\
& & \times
v\left(\left\vert\mathbf{r}-\mathbf{r'}\right\vert\right)
\psi_{j}^{*}(\mathbf{r'})\psi_{i}(\mathbf{r'})
d^{3}rd^{3}r',
\label{ExHF}
\end{eqnarray}
where the indices $i$ and $j$ run over the occupied orbitals and
$v$ is the Coulomb potential
$1/\left\vert\mathbf{r}-\mathbf{r'}\right\vert$ or only the SR
part of it\cite{BylanderPRB90,HeydJCP03} (i.e., a screened potential).
On the fifth rung of Jacob's ladder there are the functionals which utilize
all (occupied and unoccupied) orbitals, like the random phase approximation
(RPA, see, e.g., Refs. \onlinecite{HarlPRB08,OlsenPRB13}).

The functionals of the first four rungs have been extremely successful in
describing the properties of all kinds of electronic systems, ranging from atoms to
solids.\cite{CohenCR12,BurkeJCP12,BeckeJCP14} 
However, a well-known problem common to \textit{all} these functionals
is that the long-range London dispersion interactions
(always attractive and resulting from
the interaction between non-permanent multipoles) are formally not included.
In the case of two nonoverlapping spherical atoms, these functionals
give an interaction energy of strictly zero, which is not the case in reality
because of the attractive London dispersion interactions. As a consequence, the results obtained
with the semilocal and hybrid functionals on systems where the London dispersion
interactions play a major role can be qualitatively wrong.\cite{KlimesJCP12}
Nevertheless, as underlined in Ref.~\onlinecite{ZhaoJCTC05}, at equilibrium geometry
the overlap between two interacting entities is not zero, such that
a semilocal or hybrid xc-functional can eventually lead to
a non-zero contribution to the interaction energy and therefore,
possibly useful results (see, e.g., Ref.~\onlinecite{SunPRL13}).
In order to improve the reliability of KS-DFT calculations for such systems,
functionals including the dispersion interactions in their construction were proposed.
A simple and widely used method consists of adding to the semilocal or
hybrid functional an atom-pairwise (PW) term of the form
\begin{equation}
E_{\text{c,disp}}^{\text{PW}} = -\sum_{A<B}\sum_{n=6,8,10,\ldots}
f_{n}^{\text{damp}}(R_{AB})\frac{C_{n}^{AB}}{R_{AB}^{n}},
\label{Ecdisp1}
\end{equation}
where $C_{n}^{AB}$ are the dispersion coefficients for the atom pair $A$ and
$B$ separated by the distance $R_{AB}$ and $f_{n}^{\text{damp}}$ is a damping
function preventing Eq.~(\ref{Ecdisp1}) to become too large for small $R_{AB}$.
The coefficients $C_{n}^{AB}$ can be either precomputed (see, e.g., Refs.
\onlinecite{WuJCP01,WuJCP02,HasegawaPRB04,GrimmeJCC04,GrimmeJCC06,OrtmannPRB06})
or calculated using properties
of the system like the atomic positions or the electron density
(see, e.g., Refs. \onlinecite{BeckeJCP05,TkatchenkoPRL09,GrimmeJCP10}).
The other group of well-known methods accounting explicitly of dispersion
interactions consists of adding to $E_{\text{xc}}^{\text{SL/hybrid}}$
a nonlocal (NL, in the sense of being a double integral) term of the form
\cite{DionPRL04}
\begin{equation}
E_{\text{c,disp}}^{\text{NL}} = \frac{1}{2}\int\int\rho(\textbf{r})
\Phi\left(\textbf{r},\textbf{r}'\right)\rho(\textbf{r}')
d^{3}rd^{3}r',
\label{Ecdisp2}
\end{equation}
where the kernel $\Phi$ depends on $\rho$ and $\nabla\rho$ at $\textbf{r}$
and $\textbf{r}'$ as well as on $\left\vert\textbf{r}-\textbf{r}'\right\vert$.
Several functionals of the form of Eq.~(\ref{Ecdisp2}) are available in the
literature\cite{DionPRL04,LeePRB10,VydrovPRL09,VydrovJCP10,SabatiniPRB13,BerlandPRB14}
and good results can be obtained if the proper combination
$E_{\text{xc}}^{\text{SL/hybrid}}+E_{\text{c,disp}}^{\text{NL}}$ is used
(see, e.g., Refs. \onlinecite{LeePRB10,VydrovJCP10,KlimesJPCM10}).
Overall, the KS-DFT+dispersion methods produce results which are more reliable
when applied to systems where the dispersion play a major role, and
therefore, the very cheap atom-pairwise and not-too-expensive nonlocal methods
are nowadays routinely applied
(see Refs. \onlinecite{GrimmeWCMS11,KlimesJCP12,BerlandRPP15}
for recent reviews).

At this point we should certainly also mention that truly \textit{ab initio}
(beyond DFT) methods have been used for the calculation of geometrical and
energetic properties of solids (the focus of the present work). This includes
RPA, which has been shown during these last few years to be quite reliable in
many situations (see, e.g.,
Refs. \onlinecite{HarlPRB10,OlsenPRB13,SchimkaPRB13,YanPRB13} for extensive
tests), the quantum Monte Carlo methods as exemplified in
Ref.~\onlinecite{ShulenburgerPRB13} for the calculation of the lattice and bulk
modulus of a set of solids, and the post-HF methods which, as expected, should
converge to the exact results.\cite{BoothN13}

Another well-known problem in KS-DFT, that we will not address in this work, is
the inadequacy of the semilocal functionals (or more precisely of the potential
$v_{\text{xc}}=\delta E_{\text{xc}}/\delta\rho$) for the calculation of band
gaps, while the hybrid functionals work reasonably well in this respect thanks
to the nonlocal HF exchange (see, e.g.,
Refs. \onlinecite{HeydJCP05,SchimkaJCP11,PernotJPCA15}).

In the present work, a large number of functionals of
rungs 1 to 4 of Jacob's ladder, with or without a dispersion term, are
tested on solids for the calculation of lattice constant, bulk modulus, and
cohesive energy. A particular focus will be on the MGGA functionals recently
proposed by Perdew and co-workers, namely MGGA\_MS (MGGA made simple)
\cite{SunJCP12,SunJCP13,SunPRL13} and SCAN (strongly constrained and
appropriately normed semilocal density functional),\cite{SunPRL15} and
by Wellendorff \textit{et al}.,\cite{WellendorffJCP14} mBEEF
(model Bayesian error estimation functional), which should in principle
be accurate semilocal functionals for both finite and
infinite systems, and to bind systems bound by weak
interactions. Two testing sets of solids will be considered. The first one
consists of cubic elemental solids and binary compounds bound by relatively
strong interactions, while the second set is composed of systems bound mainly
by weak interactions (e.g., dispersion). This extensive study
of functionals performance on solids complements previous works, which
include Refs.
\onlinecite{TranPRB07,RopoPRB08,MattssonJCP08,CsonkaPRB09,HaasPRB09a,HaasPRB10,SunPRB11b,HaoPRB12,PeveratiPCCP12a,JanthonJCTC13,ConstantinPRB16}
for semilocal functionals, Refs.
\onlinecite{HeydJCP05,SchimkaJCP11,LuceroJPCM12,PeveratiPCCP12b,JanthonJCTC14,PernotJPCA15,RasanderJCP15} for tests
including hybrid functionals, Refs. \onlinecite{HarlPRB10,SchimkaPRB13,OlsenPRB13} for
RPA, and Refs.
\onlinecite{KlimesPRB11,BjorkmanJPCM12,BjorkmanPRB12,BjorkmanJCP14,WellendorffPRB12,WellendorffJCP14,ParkCAP15}
for a focus on functionals including dispersion via an atom-pairwise term or
a nonlocal term.

The paper is organized as follows. The computational details are given in
Sec.~\ref{details}. In Sec.~\ref{functionals}, the tested functionals are
presented and some of their features are discussed. The results are presented
and discussed in Sec.~\ref{results}, while Sec.~\ref{literature} gives a brief
literature overview of the accuracy of functionals for the energetics of
molecules. Finally, Sec.~\ref{summary} gives a summary of this work.

\section{\label{details}Computational details}

\begin{table*}
\caption{\label{solids}The test set of 44 strongly and 5 weakly bound solids
considered in this work. The space group is indicated in parenthesis.
With the exception of hexagonal graphite and h-BN, all solids are cubic.}
\begin{ruledtabular}
\begin{tabular}{l}
Strongly bound solids \\
C ($Fd\overline{3}m$), Si ($Fd\overline{3}m$), Ge ($Fd\overline{3}m$), Sn ($Fd\overline{3}m$) \\
SiC ($F\overline{4}3m$), BN ($F\overline{4}3m$), BP ($F\overline{4}3m$), AlN ($F\overline{4}3m$), AlP ($F\overline{4}3m$), AlAs ($F\overline{4}3m$) \\
GaN ($F\overline{4}3m$), GaP($F\overline{4}3m$), GaAs ($F\overline{4}3m$), InP ($F\overline{4}3m$), InAs ($F\overline{4}3m$), InSb ($F\overline{4}3m$) \\
LiH ($Fm\overline{3}m$), LiF ($Fm\overline{3}m$), LiCl ($Fm\overline{3}m$), NaF ($Fm\overline{3}m$), NaCl ($Fm\overline{3}m$), MgO ($Fm\overline{3}m$) \\
Li ($Im\overline{3}m$), Na ($Im\overline{3}m$), Al ($Fm\overline{3}m$), K ($Im\overline{3}m$), Ca ($Fm\overline{3}m$), Rb ($Im\overline{3}m$), Sr ($Fm\overline{3}m$), Cs ($Im\overline{3}m$), Ba ($Im\overline{3}m$) \\
V ($Im\overline{3}m$), Ni ($Fm\overline{3}m$), Cu ($Fm\overline{3}m$), Nb ($Im\overline{3}m$), Mo ($Im\overline{3}m$), Rh ($Fm\overline{3}m$), Pd ($Fm\overline{3}m$), Ag ($Fm\overline{3}m$) \\
Ta ($Im\overline{3}m$), W ($Im\overline{3}m$), Ir ($Fm\overline{3}m$), Pt ($Fm\overline{3}m$), Au ($Fm\overline{3}m$) \\
\hline
Weakly bound solids \\
Ne ($Fm\overline{3}m$), Ar ($Fm\overline{3}m$), Kr ($Fm\overline{3}m$), graphite ($P6_{3}/mmc$), h-BN ($P6_{3}/mmc$)
\end{tabular}
\end{ruledtabular}
\end{table*}

All calculations were done with the WIEN2k code,\cite{WIEN2k} which uses the
full-potential (linearized) augmented plane-wave plus local orbitals method
\cite{Singh} to solve the KS equations. The parameters of the calculations like
the number of $\mathbf{k}$-points for the integration of the Brillouin zone or
size of the basis set were chosen to be large enough such that the results
are well converged.

In order to make the testing of the numerous functionals tractable (especially
for the hybrids which use the expensive HF exchange), the results
on the strongly bound solids (listed in Table~\ref{solids}) were obtained
non-self-consistently by using the PBE\cite{PerdewPRL96} orbitals and densities.
According to tests, the error
in the lattice constant induced by this non-self-consistent procedure should be
in most cases below 0.005~\AA. The worst cases are the very heavy alkali and
alkali-earth metals (Cs in particular) for which the error can be of the order
of $\sim0.02$~\AA. Errors in the range 0.005-0.015~\AA~can eventually be
obtained in the case of metals with hybrid functionals
(a comparison can be done with our self-consistent hybrid
calculations reported in Refs. \onlinecite{TranPRB11,TranPRB12}).
For the cohesive energy the effect should not exceed 0.05 eV/atom except in the
eventual cases where self-consistency would lead to an atomic electronic
configuration for the isolated atom that is different from the one obtained
with PBE. For the very weakly bound rare-gas solids Ne, Ar, and Kr and
layered solids graphite and h-BN we observed
that self-consistency may have a larger impact on the results
(up to a few 0.1~\AA~in the case of very shallow total-energy curves),
therefore the calculations were done self-consistently for LDA/GGA, but not for
the MGGA functionals (not implemented self-consistently in WIEN2k) as well
as the very expensive hybrid functionals.

The results of our calculations for the strongly bound solids were compared
with experimental results that were corrected for thermal and zero-point
vibrational effects (see Refs. \onlinecite{SchimkaJCP11,LejaeghereCRSSMS14}).
For the weakly bound systems, the results were compared with accurate
\textit{ab initio} results; coupled cluster with singlet, doublet,
and perturbative triplet [CCSD(T)] for the rare gases\cite{RosciszewskiPRB00}
and RPA for the layered solids.\cite{LebeguePRL10,BjorkmanPRL12}

At this point we should remind that in general, the observed trends in
the relative performance of the functionals may depend on the test set
and more particularly on the diversity of solids.
Since our test set contains elements from all parts of the periodic
table except lanthanides and actinides, our results should give a rather fair
and unbiased overview of the accuracy of the functionals.

Finally, we mention that we did not include ferromagnetic bcc Fe in our test
set of solids, since the total-energy curves exhibit a discontinuity at the
lattice constant of $\sim2.94$~\AA~(same value as found in
Ref.~\onlinecite{SchimkaPRB13}), that is due to a change in orbitals occupation
with PBE. This discontinuity is very large when the HF exchange is used, making
an unambiguous determination of the equilibrium volume not possible with some
of the hybrid functionals.

\section{\label{functionals}The functionals}

\begingroup
\squeezetable
\begin{table*}
\caption{\label{results_strong}The ME, MAE, MRE, and MARE on the testing set of 44
strongly bound solids for the lattice constant $a_{0}$, bulk modulus $B_{0}$, and
cohesive energy $E_{\text{coh}}$. The units of the ME and MAE are \AA, GPa, and
eV/atom for $a_{0}$, $B_{0}$, and $E_{\text{coh}}$, respectively, and \% for the
MRE and MARE. All results were obtained non-self-consistently using PBE
orbitals/density. Within each group, the functionals are ordered by increasing
value of the MARE of $a_{0}$. For hybrid functionals, the fraction
$\alpha_{\text{x}}$ of HF exchange is indicated in parenthesis.}
{\tiny
\begin{ruledtabular}
\begin{tabular}{ldddddddddddd}
\multicolumn{1}{l}{} &
\multicolumn{4}{c}{$a_ {0}$} &
\multicolumn{4}{c}{$B_ {0}$} &
\multicolumn{4}{c}{$E_{\text{coh}}$} \\
\cline{2-5}\cline{6-9}\cline{10-13}
\multicolumn{1}{l}{Functional} &
\multicolumn{1}{c}{ME} &
\multicolumn{1}{c}{MAE} &
\multicolumn{1}{c}{MRE} &
\multicolumn{1}{c}{MARE} &
\multicolumn{1}{c}{ME} &
\multicolumn{1}{c}{MAE} &
\multicolumn{1}{c}{MRE} &
\multicolumn{1}{c}{MARE} &
\multicolumn{1}{c}{ME} &
\multicolumn{1}{c}{MAE} &
\multicolumn{1}{c}{MRE} &
\multicolumn{1}{c}{MARE} \\
\hline
LDA           \\
LDA \cite{PerdewPRB92a} &    -0.071 &     0.071 &      -1.5 &       1.5 &      10.1 &      11.5 &       8.1 &       9.4 &      0.77 &      0.77 &      17.2 &      17.2 \\
\hline
GGA           \\
SG4 \cite{ConstantinPRB16} &     0.005 &     0.026 &       0.0 &       0.6 &       1.7 &       7.9 &      -2.2 &       7.8 &      0.19 &      0.28 &       3.5 &       7.0 \\
WC \cite{WuPRB06} &     0.002 &     0.029 &       0.0 &       0.6 &      -0.2 &       7.6 &      -2.6 &       7.4 &      0.22 &      0.26 &       4.2 &       6.2 \\
SOGGA \cite{ZhaoJCP08} &    -0.012 &     0.027 &      -0.3 &       0.6 &       4.1 &       8.9 &       0.6 &       7.4 &      0.39 &      0.41 &       8.8 &       9.2 \\
PBEsol \cite{PerdewPRL08} &    -0.005 &     0.030 &      -0.1 &       0.6 &       0.7 &       7.8 &      -1.4 &       7.0 &      0.29 &      0.31 &       6.1 &       6.9 \\
AM05 \cite{ArmientoPRB05} &     0.014 &     0.037 &       0.2 &       0.8 &      -0.3 &       8.8 &      -4.0 &       9.2 &      0.30 &      0.45 &       7.6 &      12.6 \\
PBEint \cite{FabianoPRB10} &     0.026 &     0.039 &       0.5 &       0.8 &      -3.0 &       8.4 &      -5.3 &       8.7 &      0.10 &      0.20 &       1.5 &       4.7 \\
PBEalpha \cite{MadsenPRB07} &     0.021 &     0.042 &       0.4 &       0.9 &      -6.0 &       8.4 &      -5.0 &       7.6 &      0.10 &      0.18 &       1.8 &       4.1 \\
RGE2 \cite{RuzsinszkyJCTC09} &     0.043 &     0.051 &       0.8 &       1.0 &      -4.3 &       9.0 &      -7.3 &      10.2 &     -0.00 &      0.20 &      -1.2 &       5.0 \\
PW91 \cite{PerdewPRB92b} &     0.053 &     0.059 &       1.1 &       1.2 &     -11.0 &      12.1 &      -9.8 &      10.9 &     -0.12 &      0.18 &      -3.5 &       4.6 \\
PBE \cite{PerdewPRL96} &     0.056 &     0.061 &       1.1 &       1.2 &     -11.2 &      12.2 &      -9.8 &      11.0 &     -0.13 &      0.19 &      -3.9 &       5.0 \\
HTBS \cite{HaasPRB11} &     0.068 &     0.077 &       1.3 &       1.6 &      -4.0 &       9.9 &      -9.4 &      12.7 &     -0.14 &      0.23 &      -4.5 &       6.2 \\
PBEfe \cite{SarmientoPerezJCTC15} &     0.002 &     0.082 &       0.1 &       1.7 &     -10.0 &      12.6 &      -3.3 &      11.2 &      0.15 &      0.22 &       3.4 &       5.0 \\
revPBE \cite{ZhangPRL98} &     0.106 &     0.107 &       2.2 &       2.2 &     -17.1 &      17.5 &     -16.0 &      16.4 &     -0.48 &      0.48 &     -12.6 &      12.6 \\
RPBE \cite{HammerPRB99} &     0.119 &     0.119 &       2.4 &       2.4 &     -19.0 &      19.3 &     -17.2 &      17.5 &     -0.52 &      0.52 &     -13.2 &      13.2 \\
BLYP \cite{BeckePRA88,LeePRB88} &     0.118 &     0.120 &       2.5 &       2.5 &     -25.1 &      25.2 &     -19.9 &      20.3 &     -0.69 &      0.69 &     -20.3 &      20.3 \\
\hline
MGGA          \\
MGGA\_MS2 \cite{SunJCP13} &     0.016 &     0.029 &       0.2 &       0.6 &       4.1 &       7.6 &       0.2 &       6.8 &      0.06 &      0.21 &       0.9 &       5.2 \\
SCAN \cite{SunPRL15} &     0.018 &     0.030 &       0.3 &       0.6 &       3.5 &       7.4 &      -0.4 &       6.5 &     -0.02 &      0.19 &      -0.7 &       4.9 \\
revTPSS \cite{PerdewPRL09} &     0.023 &     0.039 &       0.4 &       0.8 &      -0.1 &       9.6 &      -3.4 &       9.4 &      0.05 &      0.22 &       1.2 &       5.1 \\
MGGA\_MS0 \cite{SunJCP12} &     0.032 &     0.044 &       0.5 &       0.9 &       4.2 &       8.3 &      -0.9 &       7.3 &     -0.02 &      0.22 &      -1.2 &       5.2 \\
MVS \cite{SunPNAS15} &    -0.008 &     0.043 &      -0.3 &       0.9 &      12.3 &      13.3 &       8.2 &      12.7 &      0.21 &      0.37 &       5.8 &       9.3 \\
mBEEF \cite{WellendorffJCP14} &     0.033 &     0.050 &       0.5 &       1.0 &       1.4 &       7.9 &      -1.4 &       7.8 &     -0.13 &      0.21 &      -3.0 &       4.6 \\
MGGA\_MS1 \cite{SunJCP13} &     0.045 &     0.054 &       0.8 &       1.1 &       1.8 &       8.1 &      -3.1 &       8.0 &     -0.10 &      0.24 &      -3.3 &       5.9 \\
TPSS \cite{TaoPRL03} &     0.045 &     0.054 &       0.9 &       1.1 &      -4.6 &       9.6 &      -6.7 &      10.3 &     -0.09 &      0.20 &      -2.3 &       4.9 \\
PKZB \cite{PerdewPRL99} &     0.086 &     0.088 &       1.7 &       1.8 &      -8.1 &      11.0 &     -10.0 &      12.4 &     -0.30 &      0.34 &      -7.2 &       8.1 \\
\hline
hybrid-LDA    \\
LDA0 \cite{PerdewPRB92a,PerdewJCP96} (0.25) &    -0.036 &     0.037 &      -0.8 &       0.9 &      12.0 &      12.6 &       7.2 &       8.7 &      0.03 &      0.31 &       0.1 &       7.5 \\
YSLDA0 \cite{PerdewPRB92a,PerdewJCP96} (0.25) &    -0.041 &     0.041 &      -0.9 &       0.9 &      11.7 &      12.2 &       7.2 &       8.6 &      0.16 &      0.30 &       3.6 &       6.8 \\
\hline
hybrid-GGA    \\
YSPBEsol0 \cite{SchimkaJCP11} (0.25) &     0.002 &     0.021 &      -0.0 &       0.5 &       6.9 &       8.7 &       1.5 &       6.9 &     -0.17 &      0.27 &      -4.0 &       6.0 \\
PBEsol0 \cite{PerdewPRL08,PerdewJCP96} (0.25) &    -0.011 &     0.021 &      -0.3 &       0.5 &      10.3 &      11.1 &       4.4 &       7.9 &     -0.13 &      0.28 &      -3.3 &       6.5 \\
PBE0 \cite{ErnzerhofJCP99,AdamoJCP99} (0.25) &     0.032 &     0.038 &       0.6 &       0.8 &       1.7 &       7.5 &      -1.8 &       6.9 &     -0.45 &      0.46 &     -10.7 &      10.9 \\
B3PW91 \cite{BeckeJCP93} (0.20) &     0.047 &     0.050 &       0.9 &       1.0 &      -3.0 &       7.5 &      -5.8 &       8.3 &     -0.55 &      0.55 &     -14.1 &      14.1 \\
YSPBE0 \cite{KrukauJCP06,TranPRB11} (0.25) &     0.054 &     0.056 &       1.0 &       1.1 &      -3.3 &       8.0 &      -5.9 &       8.6 &     -0.55 &      0.55 &     -12.9 &      12.9 \\
B3LYP \cite{StephensJPC94} (0.20) &     0.082 &     0.084 &       1.7 &       1.7 &     -13.3 &      14.5 &     -12.2 &      13.1 &     -0.84 &      0.84 &     -22.9 &      22.9 \\
\hline
hybrid-MGGA   \\
MGGA\_MS2h \cite{SunJCP13} (0.09) &     0.012 &     0.027 &       0.1 &       0.6 &       7.4 &       8.8 &       2.2 &       7.1 &     -0.07 &      0.21 &      -2.0 &       5.1 \\
revTPSSh \cite{CsonkaJCTC10} (0.10) &     0.018 &     0.033 &       0.3 &       0.7 &       3.8 &       8.8 &      -1.0 &       8.1 &     -0.09 &      0.18 &      -2.0 &       4.1 \\
TPSS0 \cite{TaoPRL03,PerdewJCP96} (0.25) &     0.025 &     0.039 &       0.4 &       0.8 &       6.6 &       9.2 &       0.5 &       8.1 &     -0.41 &      0.42 &      -9.3 &       9.4 \\
TPSSh \cite{StaroverovJCP03} (0.10) &     0.037 &     0.044 &       0.7 &       0.9 &      -0.1 &       7.7 &      -3.8 &       8.4 &     -0.22 &      0.25 &      -5.2 &       5.7 \\
MVSh \cite{SunPNAS15} (0.25) &    -0.013 &     0.055 &      -0.4 &       1.2 &      19.5 &      20.3 &      11.8 &      16.2 &     -0.19 &      0.39 &      -3.4 &       8.7 \\
\hline
GGA+D         \\
PBEsol-D3 \cite{GoerigkPCCP11} &    -0.031 &     0.039 &      -0.7 &       0.9 &       5.9 &      10.0 &       2.7 &       7.3 &      0.50 &      0.50 &      11.7 &      11.7 \\
PBE-D3 \cite{GrimmeJCP10} &     0.022 &     0.042 &       0.4 &       0.9 &      -4.8 &       8.7 &      -5.0 &       8.3 &      0.12 &      0.16 &       2.9 &       3.9 \\
PBE-D3(BJ) \cite{GrimmeJCC11} &    -0.002 &     0.042 &      -0.1 &       0.9 &      -3.1 &       7.5 &      -2.1 &       7.4 &      0.20 &      0.21 &       4.8 &       5.2 \\
revPBE-D3(BJ) \cite{GrimmeJCC11} &    -0.011 &     0.043 &      -0.4 &       1.0 &      -0.4 &       8.5 &      -1.4 &       8.6 &      0.18 &      0.21 &       4.2 &       5.2 \\
revPBE-D3 \cite{GrimmeJCP10} &     0.042 &     0.060 &       0.7 &       1.2 &      -6.9 &      11.9 &      -7.4 &      13.4 &     -0.02 &      0.18 &      -0.3 &       4.4 \\
PBEsol-D3(BJ) \cite{GoerigkPCCP11} &    -0.060 &     0.061 &      -1.3 &       1.3 &       8.6 &      11.2 &       6.2 &       8.7 &      0.62 &      0.62 &      14.9 &      14.9 \\
RPBE-D3 \cite{DFTD3} &     0.063 &     0.070 &       1.2 &       1.4 &     -13.9 &      15.2 &     -10.6 &      14.1 &     -0.14 &      0.20 &      -2.9 &       5.0 \\
BLYP-D3 \cite{GrimmeJCP10} &     0.043 &     0.070 &       0.7 &       1.4 &     -12.3 &      16.1 &     -10.5 &      15.6 &     -0.18 &      0.24 &      -6.6 &       7.7 \\
BLYP-D3(BJ) \cite{GrimmeJCC11} &    -0.034 &     0.074 &      -0.8 &       1.6 &      -5.7 &      10.8 &      -1.0 &      10.6 &      0.11 &      0.21 &       0.8 &       6.1 \\
\hline
MGGA+D        \\
MGGA\_MS2-D3 \cite{SunJCP13} &     0.002 &     0.030 &      -0.1 &       0.6 &       7.5 &       9.9 &       2.7 &       8.1 &      0.17 &      0.25 &       3.7 &       5.8 \\
MGGA\_MS0-D3 \cite{SunJCP13} &     0.019 &     0.040 &       0.2 &       0.8 &       7.6 &      10.5 &       1.5 &       8.9 &      0.08 &      0.21 &       1.5 &       4.9 \\
MGGA\_MS1-D3 \cite{SunJCP13} &     0.026 &     0.047 &       0.4 &       1.0 &       6.4 &      10.5 &       0.2 &       8.9 &      0.05 &      0.20 &       0.5 &       4.6 \\
TPSS-D3 \cite{GrimmeJCP10} &    -0.004 &     0.045 &      -0.3 &       1.0 &       5.9 &      14.0 &       0.9 &      11.2 &      0.27 &      0.30 &       7.5 &       8.1 \\
TPSS-D3(BJ) \cite{GrimmeJCC11} &    -0.042 &     0.049 &      -1.0 &       1.1 &       8.7 &      12.9 &       5.1 &      10.3 &      0.42 &      0.42 &      11.0 &      11.1 \\
\hline
hybrid-GGA+D  \\
PBE0-D3 \cite{GrimmeJCP10} (0.25) &    -0.005 &     0.027 &      -0.3 &       0.6 &       9.3 &      11.7 &       4.0 &       9.0 &     -0.16 &      0.23 &      -2.9 &       4.8 \\
YSPBE0-D3(BJ) \cite{DFTD3} (0.25) &    -0.023 &     0.030 &      -0.6 &       0.7 &       7.8 &       9.9 &       4.4 &       7.7 &     -0.10 &      0.22 &      -0.9 &       5.2 \\
PBE0-D3(BJ) \cite{GrimmeJCC11} (0.25) &    -0.030 &     0.035 &      -0.7 &       0.8 &      11.3 &      12.2 &       7.2 &       9.3 &     -0.07 &      0.24 &      -0.7 &       5.2 \\
YSPBE0-D3 \cite{DFTD3} (0.25) &     0.035 &     0.042 &       0.6 &       0.8 &       0.8 &       7.4 &      -3.0 &       7.5 &     -0.41 &      0.41 &      -9.3 &       9.4 \\
B3LYP-D3 \cite{GrimmeJCP10} (0.20) &     0.018 &     0.047 &       0.2 &       1.0 &      -1.3 &      10.5 &      -3.1 &      11.3 &     -0.38 &      0.38 &     -10.2 &      10.4 \\
B3LYP-D3(BJ) \cite{GrimmeJCC11} (0.20) &    -0.043 &     0.055 &      -1.0 &       1.2 &       4.0 &       8.7 &       4.9 &       8.7 &     -0.15 &      0.22 &      -4.6 &       6.4 \\
\hline
hybrid-MGGA+D \\
MGGA\_MS2h-D3 \cite{SunJCP13} (0.09) &    -0.002 &     0.030 &      -0.2 &       0.7 &      10.9 &      11.7 &       4.7 &       9.1 &      0.04 &      0.21 &       0.8 &       5.1 \\
TPSSh-D3 \cite{GoerigkPCCP11} (0.10) &    -0.013 &     0.040 &      -0.5 &       0.9 &      10.6 &      14.5 &       4.1 &      11.1 &      0.16 &      0.19 &       5.0 &       5.7 \\
TPSS0-D3 \cite{GrimmeJCP10} (0.25) &    -0.023 &     0.045 &      -0.7 &       1.1 &      17.6 &      18.7 &       8.6 &      13.4 &     -0.03 &      0.20 &       0.9 &       5.1 \\
TPSSh-D3(BJ) \cite{HoffmannJCC14} (0.10) &    -0.049 &     0.054 &      -1.2 &       1.2 &      13.5 &      15.0 &       8.3 &      11.4 &      0.30 &      0.30 &       8.5 &       8.5 \\
TPSS0-D3(BJ) \cite{GrimmeJCC11} (0.25) &    -0.064 &     0.068 &      -1.5 &       1.6 &      20.9 &      21.0 &      14.0 &      15.3 &      0.13 &      0.24 &       4.9 &       7.0 \\
\end{tabular}
\end{ruledtabular}
}
\end{table*}
\endgroup

The exchange-correlation functionals that were tested for the present work are
listed in Table~\ref{results_strong}. They are grouped into families, namely,
LDA, GGA, and MGGA, and their extensions that use the HF exchange,
Eq.~(\ref{Exchybrid}), [hybrid-\ldots], a dispersion correction of the
atom-pairwise type as given by Eq.~(\ref{Ecdisp1}) [\ldots+D], or both.
Among the GGA functionals, BLYP\cite{BeckePRA88,LeePRB88} and
PBE\cite{PerdewPRL96} have been the most used in chemistry and physics,
respectively. PBE leads to reasonable results for solids (lattice constant
and cohesive energy), while BLYP is much more appropriate for the
atomization energy of molecules. More recent GGA functionals are
AM05,\cite{ArmientoPRB05} WC,\cite{WuPRB06} SOGGA,\cite{ZhaoJCP08} and
PBEsol,\cite{PerdewPRL08} which are more accurate for the lattice constant of
solids (see, e.g., Ref.~\onlinecite{HaasPRB09a}), but severely overbind
molecules.\cite{HaasPRB11} Other recent GGA functionals that were also tested
are PBEint,\cite{FabianoPRB10} PBEfe,\cite{SarmientoPerezJCTC15} and
SG4.\cite{ConstantinPRB16} In the group of MGGA functionals,
there are the relatively old functionals PKZB\cite{PerdewPRL99} and
TPSS,\cite{TaoPRL03} as well as the very recent ones
MGGA\_MS2,\cite{SunJCP13,SunPRL13} mBEEF,\cite{WellendorffJCP14} and
SCAN,\cite{SunPRL15} which should be among the most accurate semilocal
functionals for molecules and solids and also provide (possibly) useful
results for weakly bound systems. The other recent MGGA MVS\cite{SunPNAS15}
is also among the tested functionals.

A semilocal functional can be defined by its xc-enhancement factor $F_{\text{xc}}$:
\begin{equation}
F_{\text{xc}}(\mathbf{r}) =
\frac{\epsilon_{\text{xc}}(\mathbf{r})}{\epsilon_{\text{x}}^{\text{LDA}}(\mathbf{r})} =
\frac{\epsilon_{\text{x}}(\mathbf{r})+\epsilon_{\text{c}}(\mathbf{r})}{\epsilon_{\text{x}}^{\text{LDA}}(\mathbf{r})} =
F_{\text{x}}(\mathbf{r}) + F_{\text{c}}(\mathbf{r}),
\label{Fxc}
\end{equation}
where $\epsilon_{\text{x}}^{\text{LDA}}=-\left(3/4\right)\left(3/\pi\right)^{1/3}\rho^{4/3}$
is the exact exchange-energy density for constant electron densities.
\cite{DiracPCPS30,GasparAPH54,KohnPR65}
For convenience, $F_{\text{xc}}$ is usually expressed as a function of the variables
$r_{s}=\left(3/\left(4\pi\rho\right)\right)^{1/3}$ (the radius of the sphere
which contains one electron),
$s=\left\vert\nabla\rho\right\vert/\left(2\left(3\pi^{2}\right)^{1/3}\rho^{4/3}\right)$
(the reduced density gradient), and $\alpha=\left(t-t^{\text{W}}\right)/t^{\text{TF}}$
where $t^{\text{W}}=\left\vert\nabla\rho\right\vert^{2}/\left(8\rho\right)$ is
the von Weizs\"{a}cker\cite{vonWeizsackerZP35} KE density
(exact for systems with only one occupied orbital)
and $t^{\text{TF}}=\left(3/10\right)\left(3\pi^{2}\right)^{2/3}\rho^{5/3}$ is the
Thomas-Fermi KE density\cite{ThomasPCPS27,FermiRANL27}
(exact for constant electron densities).
Note that the exchange part $F_{\text{x}}$ does not depend on $r_{s}$, but only
on $s$ (and $\alpha$ for MGGAs).
\begin{figure}
\includegraphics[scale=0.65]{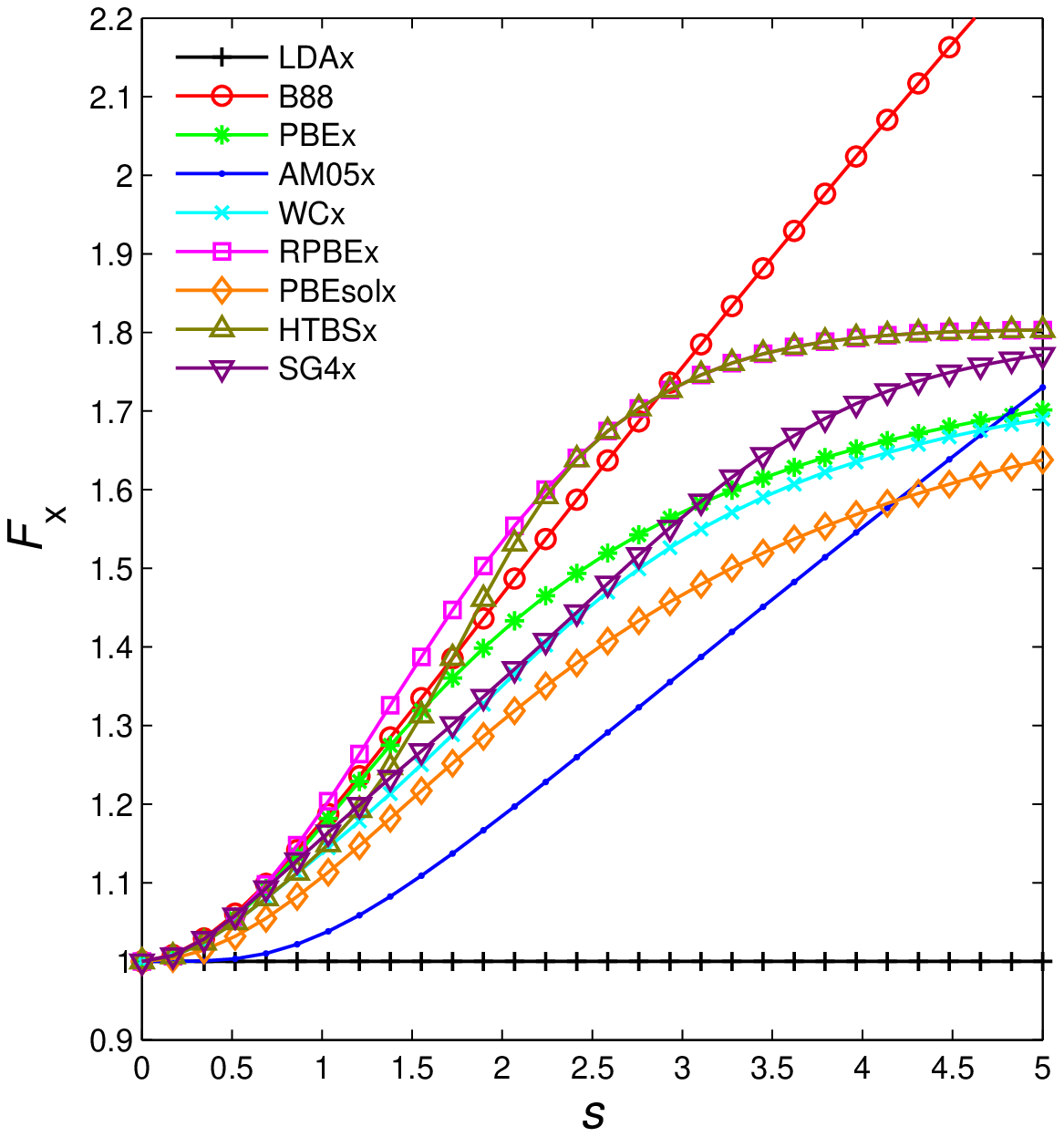}
\includegraphics[scale=0.67]{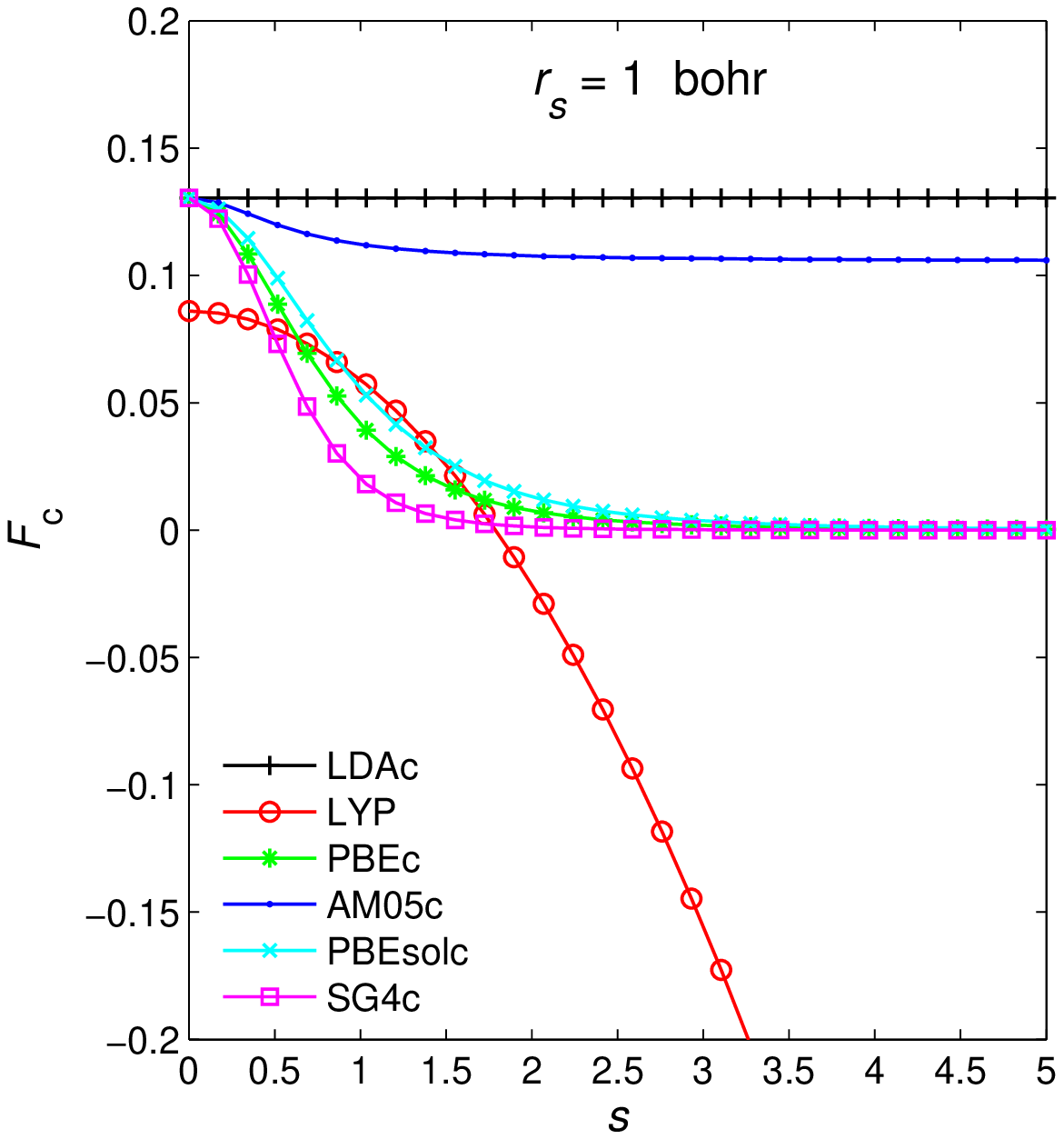}
\caption{\label{fig_Fxc1}GGA enhancement factors $F_{\text{x}}$
(upper panel) and $F_{\text{c}}$ (lower panel, for $r_{s}=1$ bohr)
plotted as a function of $s$.}
\end{figure}
\begin{figure}
\includegraphics[scale=0.35]{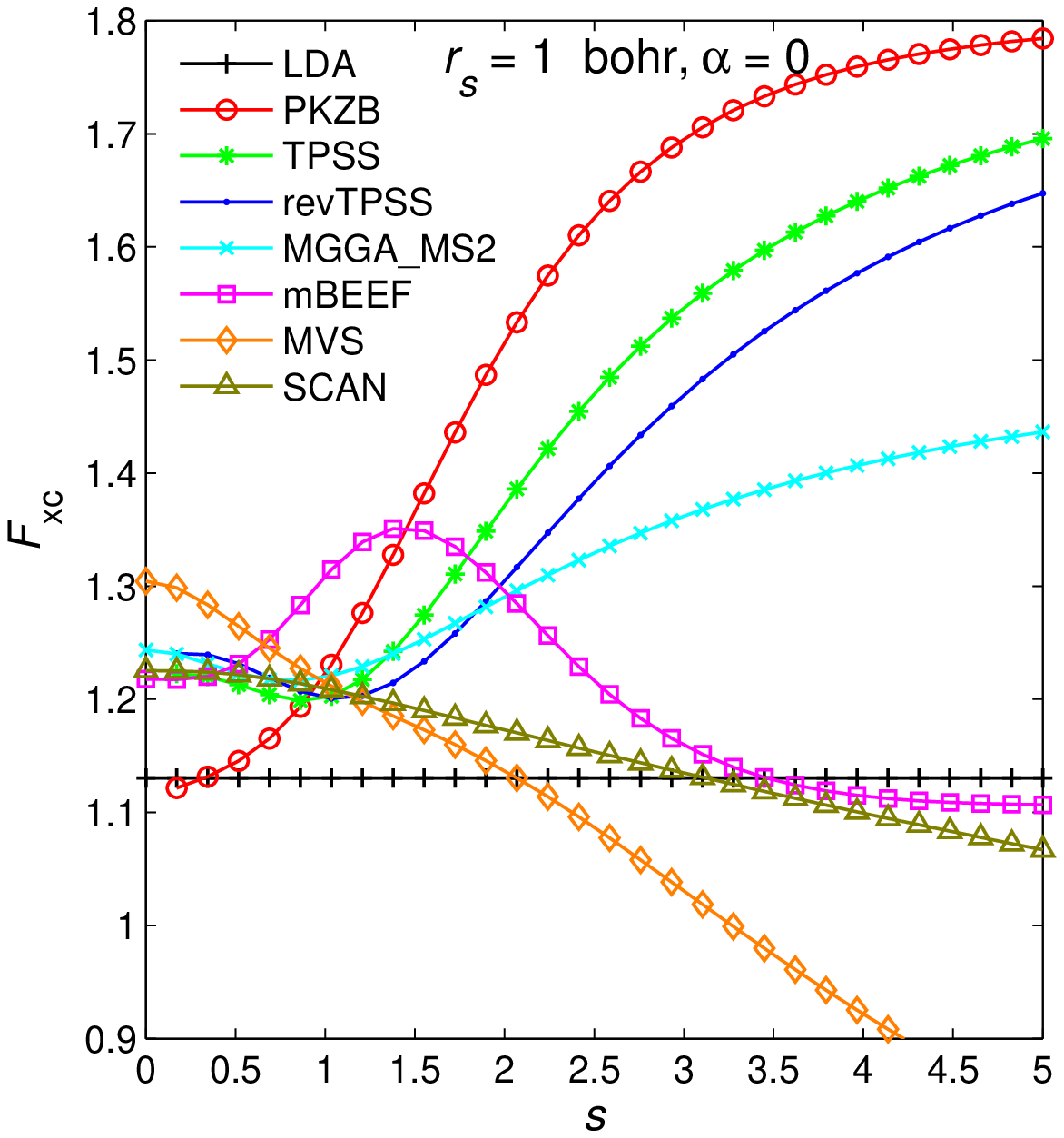}
\includegraphics[scale=0.35]{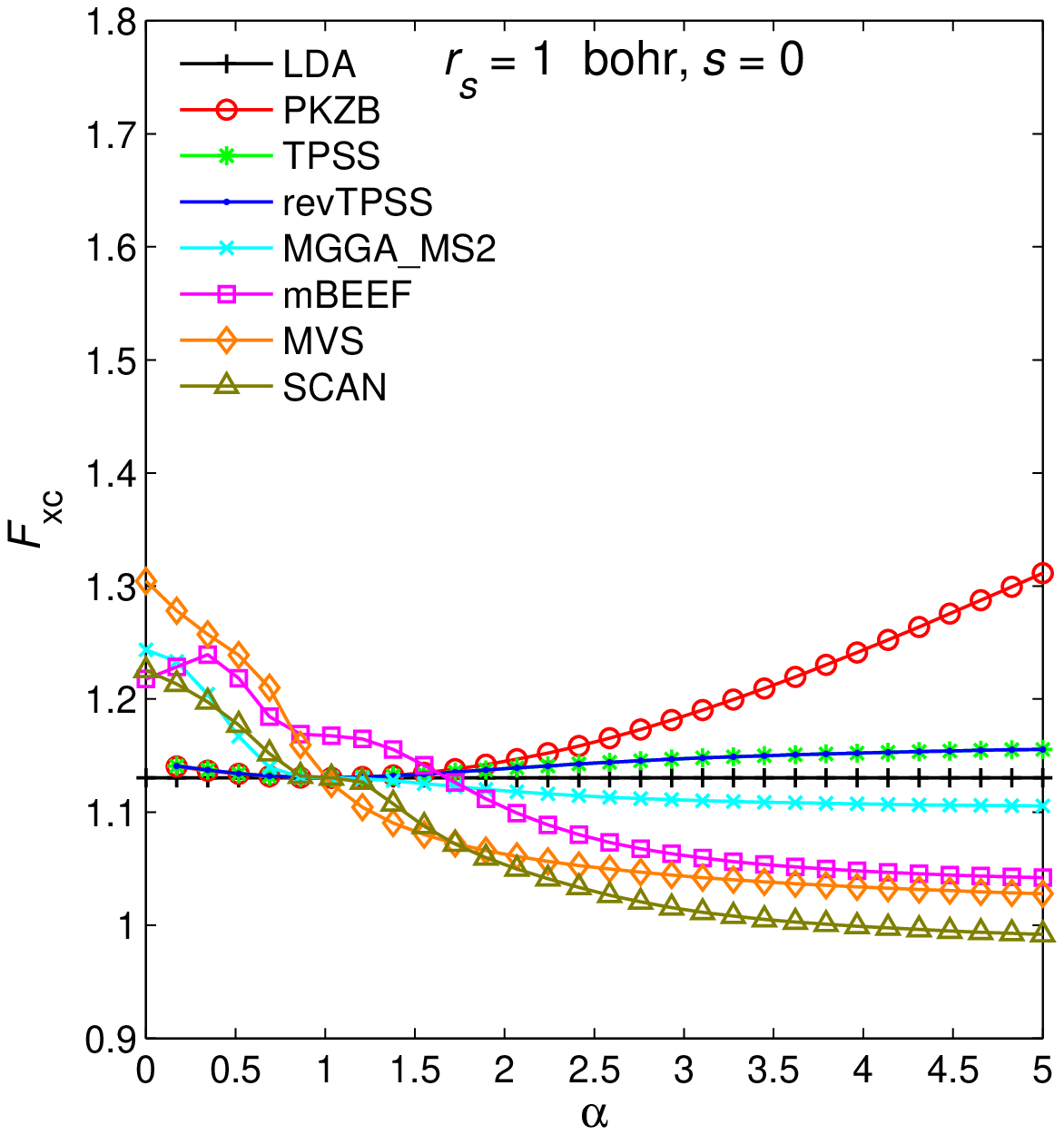}
\includegraphics[scale=0.35]{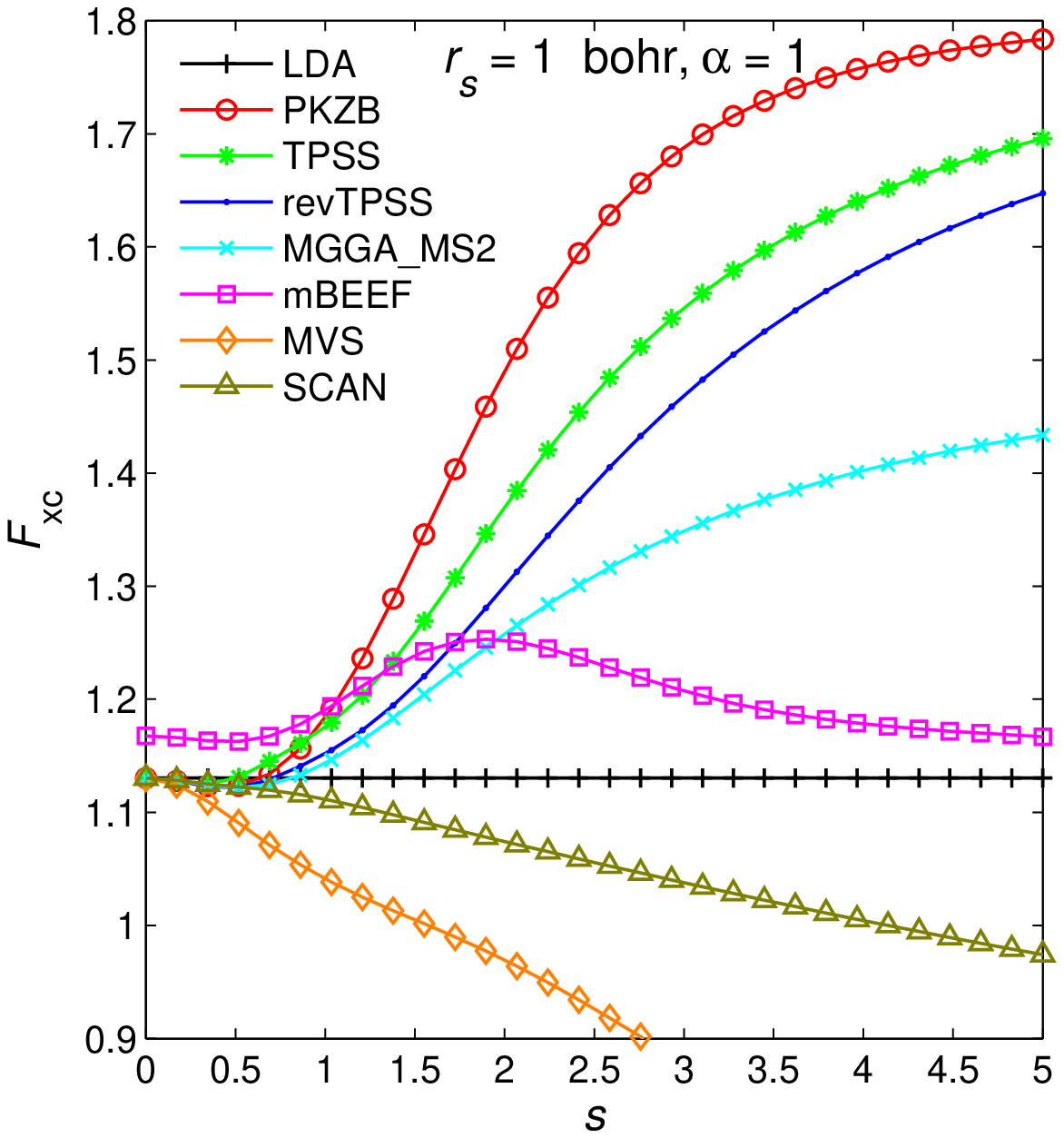}
\includegraphics[scale=0.35]{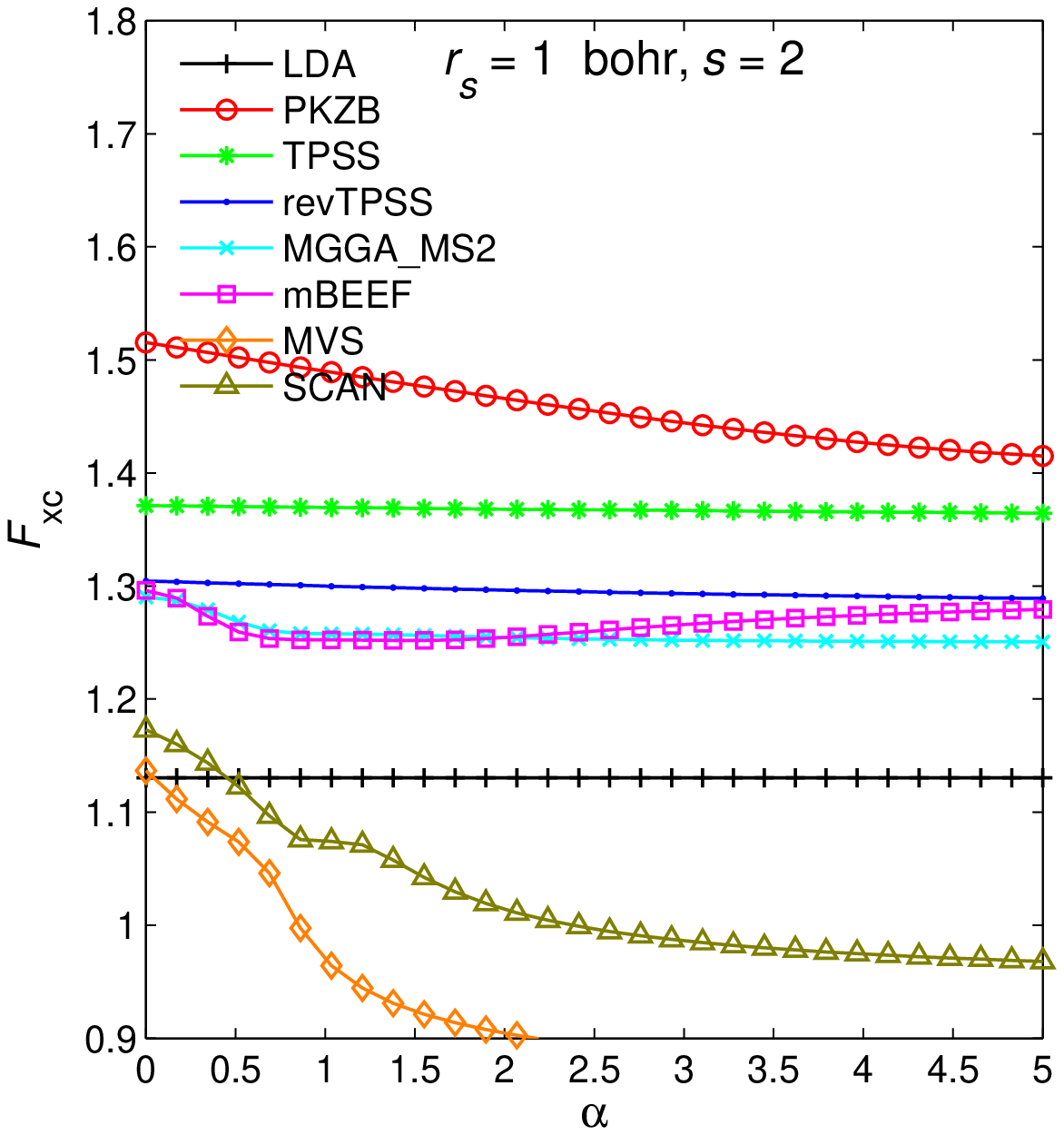}
\includegraphics[scale=0.35]{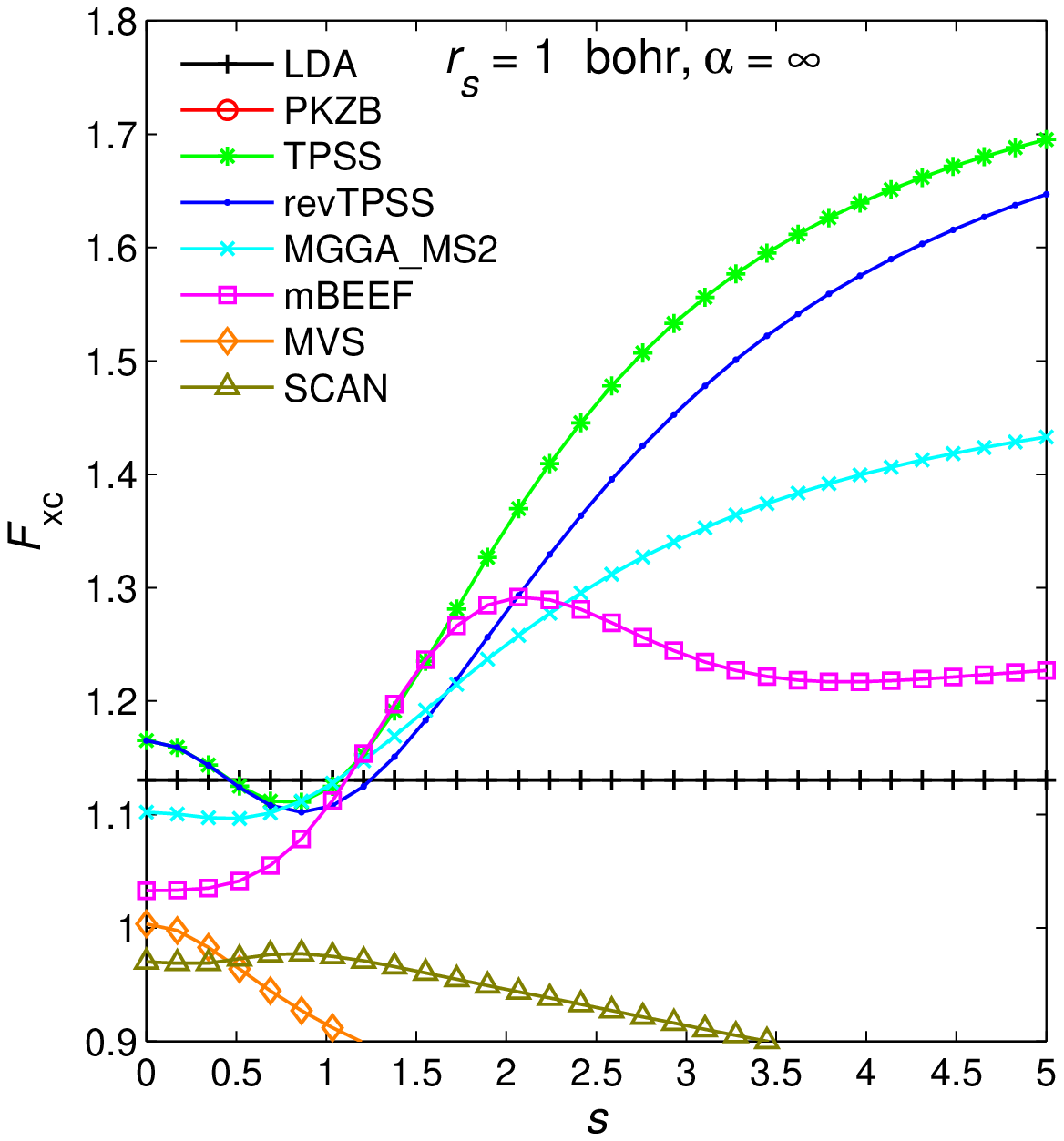}
\includegraphics[scale=0.35]{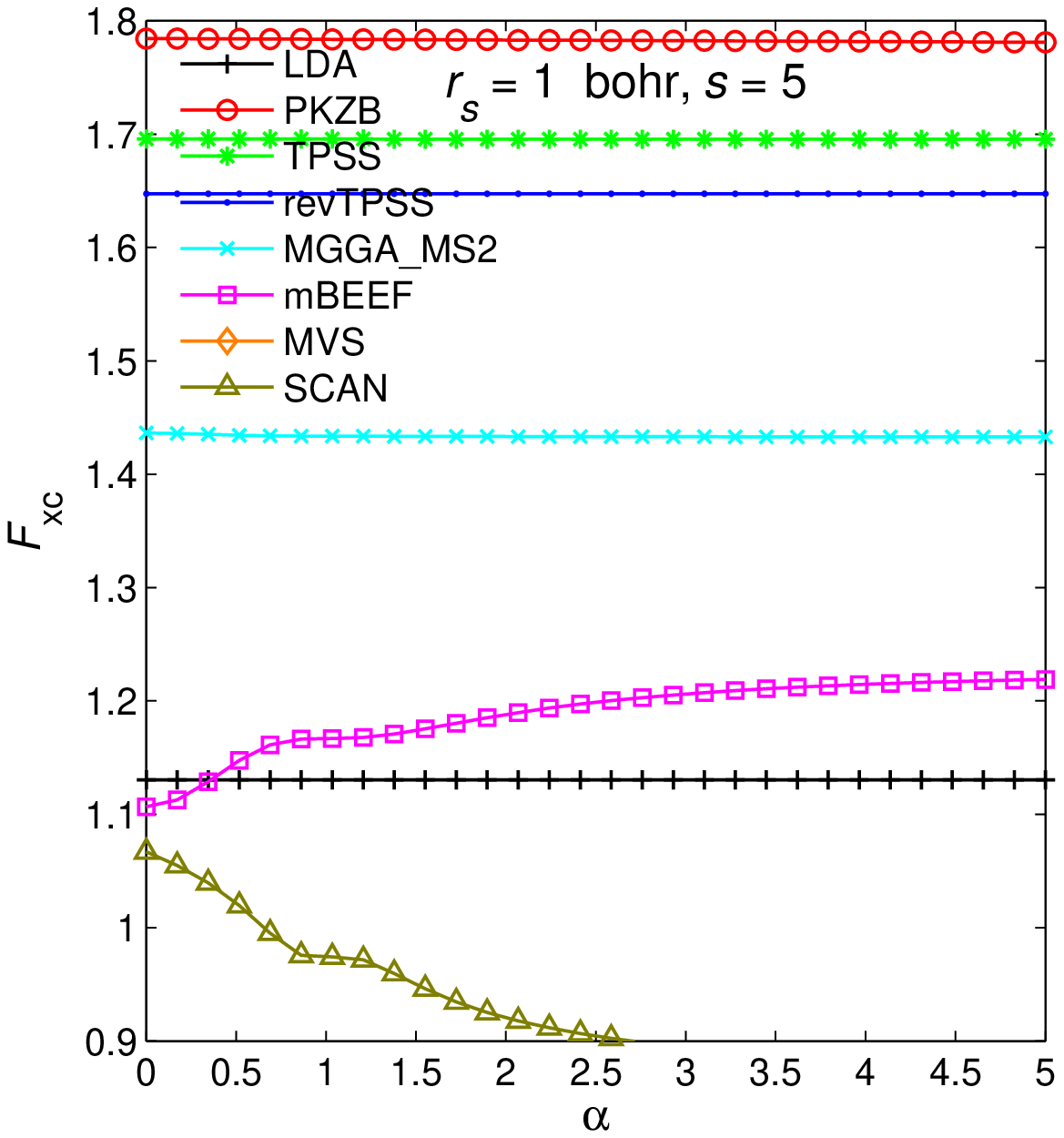}
\caption{\label{fig_Fxc2}MGGA enhancement factors $F_{\text{xc}}$
plotted as a function of $s$ for three values of
$\alpha$ (left panels) and as a function of $\alpha$ for three values
of $s$ (right panels). $r_{s}$ is kept fixed to 1 bohr.
The $F_{\text{xc}}$ for LDA is also shown.}
\end{figure}

Figures \ref{fig_Fxc1} and \ref{fig_Fxc2} show the enhancement factor of most
GGA and MGGA functionals tested in this work.
Now, we summarize the trends in the performances of GGA functionals and
how their are related to the shape of $F_{\text{xc}}$ (mainly determined by the
dominating exchange part $F_{\text{x}}$, see upper panel of Fig.~\ref{fig_Fxc1}).
LDA, which has the \textit{weakest} enhancement factor underestimates the
equilibrium lattice constant $a_{0}$ of solids. Since large unit cells contain more
density gradient (i.e., larger $s$) than small unit cells, then a
stronger $F_{\text{xc}}$ (any GGA, see Fig.~\ref{fig_Fxc1}) will lower the total
energy more for large unit cells than for smaller ones (a stronger
$F_{\text{xc}}$ makes the total energy more negative), and therefore reduce
the underestimation of $a_{0}$ obtained with LDA. A good balance is obtained with
\textit{weak} GGAs like AM05 or PBEsol (see Fig.~\ref{fig_Fxc1}) that are among the most
accurate for lattice constants. Concerning the cohesive energy $E_{\text{coh}}$ of solids, LDA
overestimates the values. Since an isolated atom contains much more density
gradient than the solid, then a GGA (w.r.t. LDA) lowers the total energy
of the atom by a larger amount than for the solid, thus reducing the
overestimation of $E_{\text{coh}}$. In this respect, functionals with a
\textit{medium} $F_{\text{xc}}$ like PBE do a pretty good job. GGAs
with a \textit{strong} $F_{\text{xc}}$ like B88 or RPBE overcorrect LDA
and lead to overestimation and underestimation
of $a_{0}$ and $E_{\text{coh}}$, respectively. LDA overestimates the
atomization energies of molecules as well, and, using the same argument as
for $E_{\text{coh}}$, a GGA lowers (w.r.t. LDA) the total energy more for the atoms
than for the molecule. However, in this case, functionals with a
\textit{strong} enhancement factor (e.g., B88) are the best performing GGAs, while
weaker GGAs reduce only partially the LDA overestimation.
One may ask the following question: Why is it necessary to use a $F_{\text{xc}}$
that is stronger for the atomization energy of molecules than for the cohesive
energy of solids? The reason is that the degrees of $\rho$-inhomogeneity in the solid
and atom (both used to calculate $E_{\text{coh}}$) are very different, such
that the appropriate difference (between the solid and atom) in the lowering
of the total energy (w.r.t. LDA) is already achieved with a weak
$F_{\text{xc}}$. Since the atomization energy of molecules requires
calculations on the atoms and molecule, which have more similar
inhomogeneities (slightly larger in atoms than in molecule), then a stronger
$F_{\text{xc}}$ is required to achieve the desired difference (between the
atoms and the molecule) in the lowering of the total energy (w.r.t. LDA).

The above trends hold for GGA functionals that are conventional in the sense
that $F_{\text{xc}}$ does not exhibit a strange behavior like oscillations or
a suddenly large slope $\partial F_{\text{xc}}/\partial s$ in a small region of $s$.
More unconventional forms for $F_{\text{xc}}$ are usually obtained when
$F_{\text{xc}}$ contains empirical parameters that are determined by a fit of
reference data.
The problem of such functionals is a reduced degree
of transferability and clear failures in particular cases.
An example of such a functional
is given by the GGA exchange HTBS\cite{HaasPRB11} (shown in Fig.~\ref{fig_Fxc1})
that was an attempt to construct a functional which leads to good
results for both the lattice constants of solids and atomization energies of
molecules: $F_{\text{x}}^{\text{HTBS}}=F_{\text{x}}^{\text{WC}}$ for
$s<0.6$ (weak GGA for $s$ values relevant for solids) and
$F_{\text{x}}^{\text{HTBS}}=F_{\text{x}}^{\text{RPBE}}$ for $s>2.6$
(strong GGA for $s$ values relevant for finite systems), while
a linear combination of WC and RPBE is used for $s\in[0.6,2.6]$.
The results were shown to be very good except for systems containing alkali
metals whose lattice constants are largely overestimated,\cite{HaasPRB11}
which is due to the large values of $s$ in the core-valence separation region
of the alkali metals (see Ref. \onlinecite{HaasPRB09b}).

Concerning the correlation enhancement factor $F_{\text{c}}$
(see lower panel of Fig.~\ref{fig_Fxc1}), we just note that
for LYP it behaves differently from the others and that
the LDA limit is not recovered, since LYP was designed
to reproduce the correlation energy of the helium atom and
not of a constant electron density.\cite{LeePRB88}

In Fig.~\ref{fig_Fxc2}, the total enhancement factor $F_{\text{xc}}$ of MGGAs
is plotted as a function of $s$ for three values of $\alpha$
(on the left panels) and as a function of $\alpha$ for three values of $s$
(on the right panels). The three values of $\alpha$ correspond to regions
dominated by a single orbital ($\alpha=0$),
of constant electron density ($\alpha=1$), and of overlap of closed shells
($\alpha\gg5$).\cite{SunPRL13}
A few comments can be made. As mentioned above,
it can happen for parameterized functionals,
like the Minnesota suite of functionals,\cite{ZhaoTCA08}
to have an enhancement factor that shows features like bumps or oscillations
that are unphysical and may lead to problems of transferability
of the functional. Furthermore, such features lead to numerical
noise.\cite{JohnsonJCP09,MardirossianJCP15}
The mBEEF is such a highly parameterized functional, however, since
its parameters were fitted with a regularization procedure\cite{WellendorffJCP14}
the bump visible in Fig.~\ref{fig_Fxc2} is moderate. A particularity of the SCAN
and MVS enhancement factors is to be much more a decreasing function of $s$
and $\alpha$ than the other functionals. Note also the very weak variation
of $F_{\text{xc}}$ with respect to $\alpha$ for the TPSS, revTPSS, and MGGA\_MS2
functionals. For MGGA functionals, it is maybe more difficult than for GGAs to
establish simple relations between the shape of $F_{\text{xc}}$ and the
trends in the results. Anyway, it is clear that the additional
dependency on the KE density leads to more flexibility and therefore
potentially more universal functionals. Finally, we mention
Refs. \onlinecite{MadsenJPCL10,SunPRL13}, where
it was argued that at small $s$ the enhancement factor should be a decreasing
function of $\alpha$ in order to obtain a binding between weakly interacting
systems, which is the case for MGGA\_MS2, MVS, and SCAN as shown in
Fig.~\ref{fig_Fxc2}.

The hybrid functionals can be split into two groups according to the type of
HF exchange that is used: the ones that use the unscreened HF exchange and
those using only the SR part that was obtained by means of the screened Yukawa
potential (details of the implementation can be found in
Ref.~\onlinecite{TranPRB11}). The screened hybrid functionals in
Table~\ref{results_strong} are those whose name starts with YS (Yukawa screened), and
among them, YSPBE0\cite{TranPRB11} is based on the popular functional of Heyd
\textit{et al}.\cite{HeydJCP03,KrukauJCP06} HSE06 and differs from it by the
screening (Yukawa in YSPBE0 versus error function in HSE06) and the way the
screening is applied in the semilocal exchange (via the exchange
hole\cite{HeydJCP03} or via the GGA enhancement factor\cite{IikuraJCP01}).
As noticed in Ref.~\onlinecite{ShimazakiCPL08}, the error function- and
Yukawa-screened potentials are very similar if the screening parameter in the
Yukawa potential is $3/2$ larger than in the error function.
In Ref.~\onlinecite{TranPRB11} it was shown that HSE06 and YSPBE0 lead to
basically the same band gaps, while non-negligible differences were observed
for the lattice constants. A comparison between the YSPBE0 results obtained in the present
work and the HSE06 results reported in Ref.~\onlinecite{SchimkaJCP11} shows
that the YSPBE0 lattice constants are in most cases slightly larger by
0.01-0.02~\AA, while the atomization energies can differ by 0.05-0.2 eV/atom.
Similarly, YSPBEsol0 uses the same underlying semilocal functional (PBEsol\cite{PerdewPRL09})
and fraction of HF exchange (0.25) as HSEsol (Ref.~\onlinecite{SchimkaJCP11}).
For all screened hybrid functionals tested in this work, a screening parameter
$\lambda=0.165$ bohr$^{-1}$ was used, which is $3/2$ of the value used in
HSE06 with the error function.\cite{KrukauJCP06} The fraction $\alpha_{\text{x}}$
of HF exchange (indicated in Table~\ref{results_strong}) varies between 0.09
(MGGA\_MS2h) and 0.25 (e.g., PBE0).
Among the unscreened
hybrid functionals in Table~\ref{results_strong}, the two most well-known are
B3LYP\cite{BeckeJCP93,StephensJPC94} and PBE0.\cite{ErnzerhofJCP99,AdamoJCP99}
Note that in Refs. \onlinecite{TranPRB12,JangJPSJ12,Gao15}, the use of hybrid
functionals for metals has been severely criticized, since qualitatively
wrong results (e.g., incorrect prediction for the ground state or largely
overestimated magnetic moment) were obtained for simple transition metals
like Fe or Pd.

The two variants of atom-pairwise dispersion correction [Eq.~(\ref{Ecdisp1})]
that are considered
were proposed by Grimme and co-workers.\cite{GrimmeJCP10,GrimmeJCC11}
The two schemes, which use the position of atoms to calculate the dispersion
coefficients $C_{n}^{AB}$,
differ in the damping function $f_{n}^{\text{damp}}$.
In the first scheme (DFT-D3, Ref.~\onlinecite{GrimmeJCP10}), the dispersion
energy $E_{\text{c,disp}}^{\text{PW}}$ goes to zero when $R_{AB}\rightarrow0$,
while with the Becke-Johnson (BJ) damping function\cite{JohnsonJCP06} that is
used in DFT-D3(BJ),\cite{GrimmeJCC11} $E_{\text{c,disp}}^{\text{PW}}$ goes to a
nonzero value, which is theoretically correct.\cite{KoideJPB76}
All DFT-D3/D3(BJ) calculations were done with and without the three-body
non-additive dispersion term,\cite{GrimmeJCP10} which has little influence on
the results for the strongly bound and rare-gas solids. For the layered
compounds, however, the effect is larger since adding the three-body term increases the
equilibrium lattice constant $c_{0}$ by $\sim0.1$~\AA~and decreases the interlayer binding
energy by $\sim10$~meV/atom, which for the latter quantity leads to better
agreement with the references results in most cases. In the following, only the
results including the three-body term will be shown. Note that in the case of
YSPBE0-D3/D3(BJ), the parameters of the D3/D3(BJ) corrections
are those that were proposed for the HSE06 functional. The DFT-D3/D3(BJ)
dispersion energies were evaluated by using the package provided by
Grimme\cite{DFTD3} that supports periodic boundary conditions.
\cite{MoellmannJPCM12,MoellmannJPCC14}

\section{\label{results}Results and discussion}

\subsection{\label{normal}Strongly bound solids}

\begin{figure*}
\includegraphics[scale=0.56]{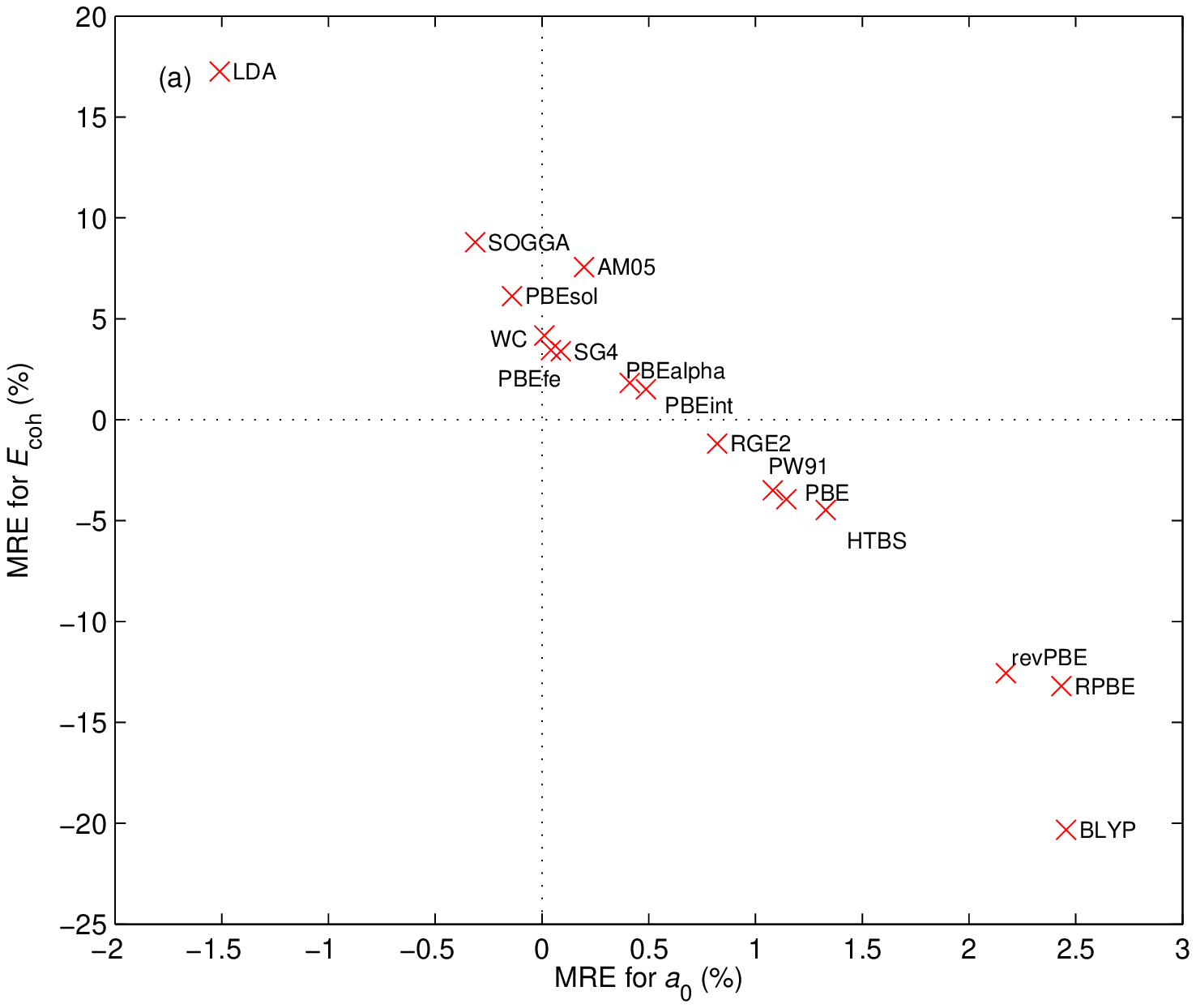}
\includegraphics[scale=0.56]{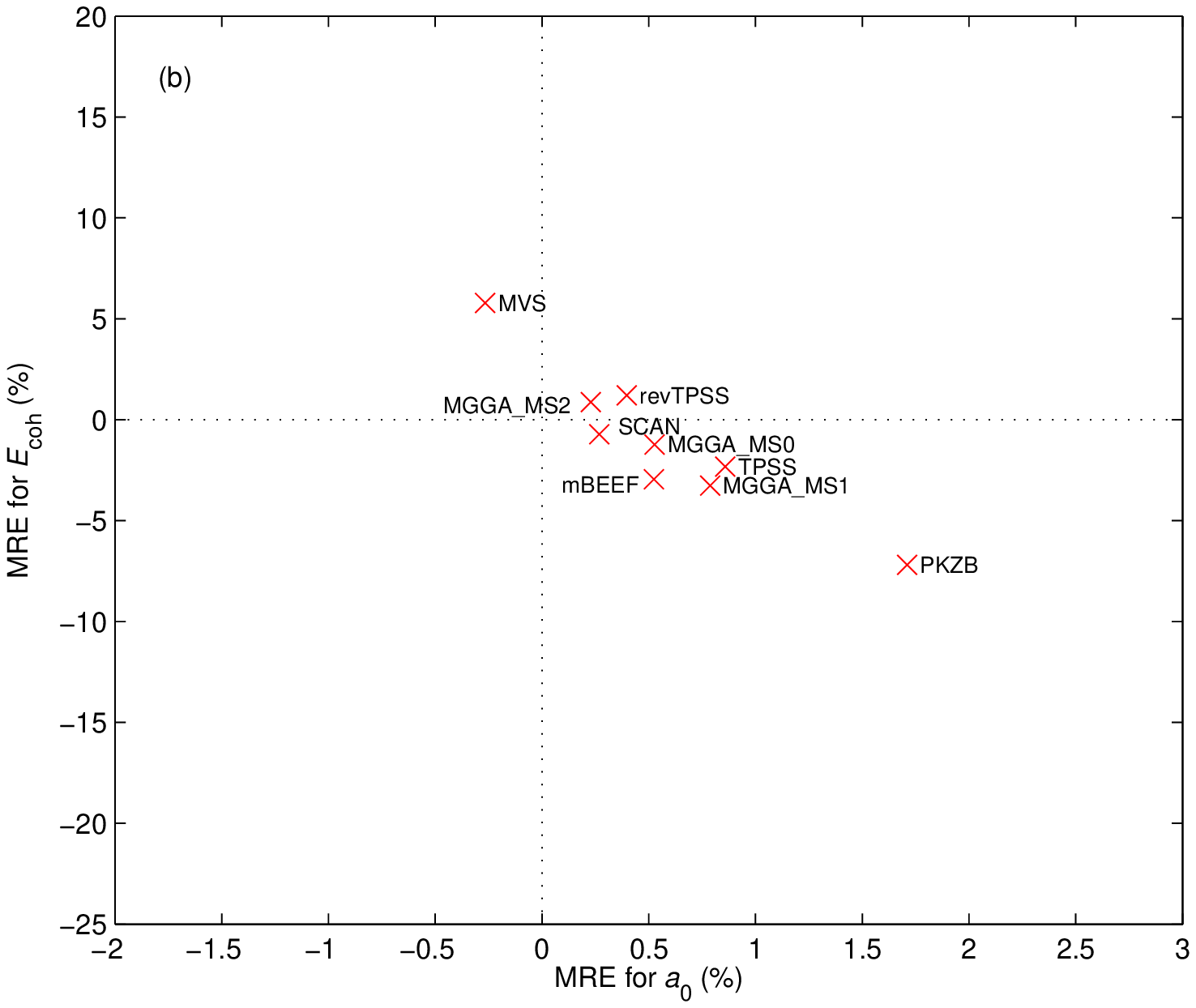}
\includegraphics[scale=0.56]{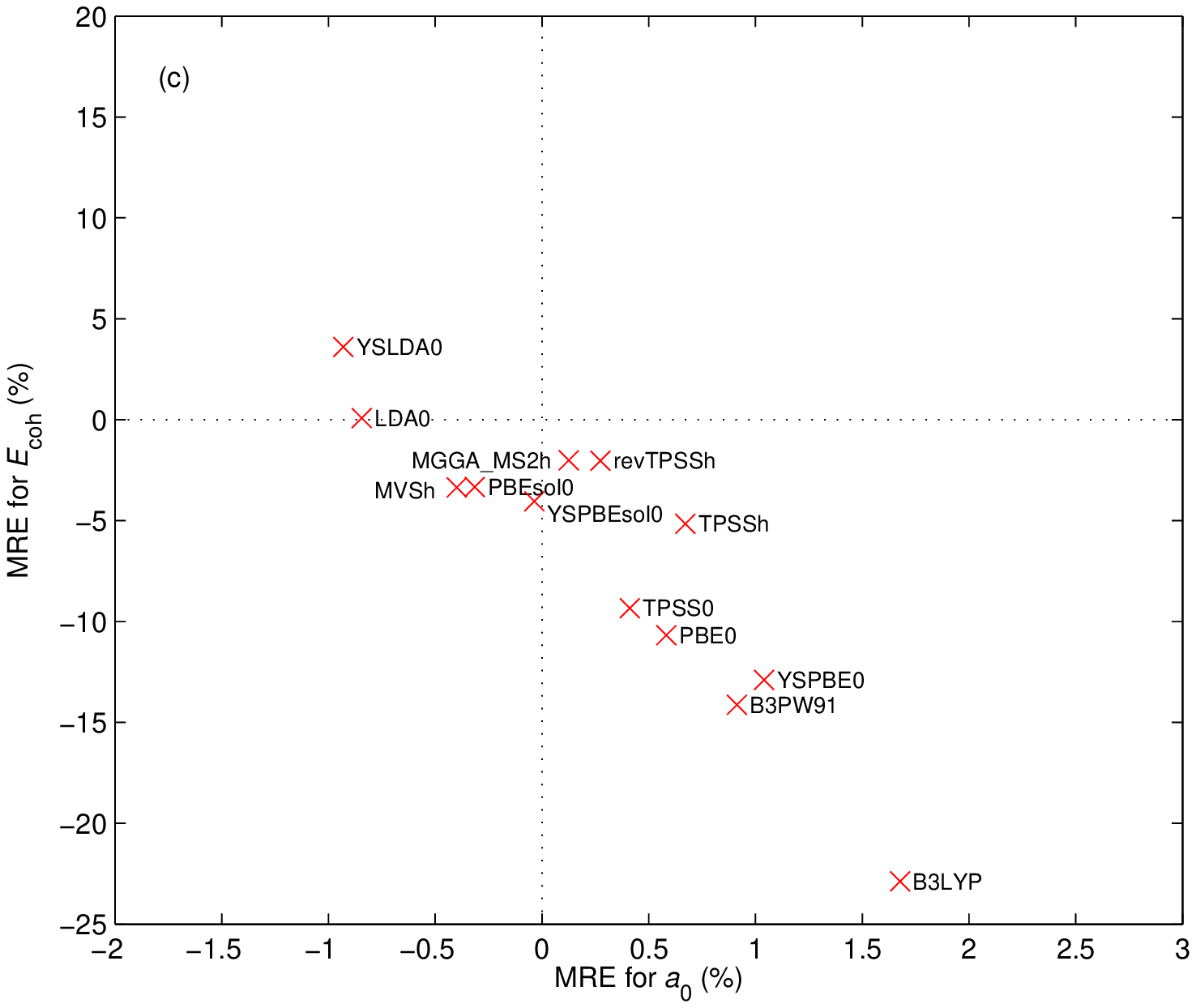}
\includegraphics[scale=0.56]{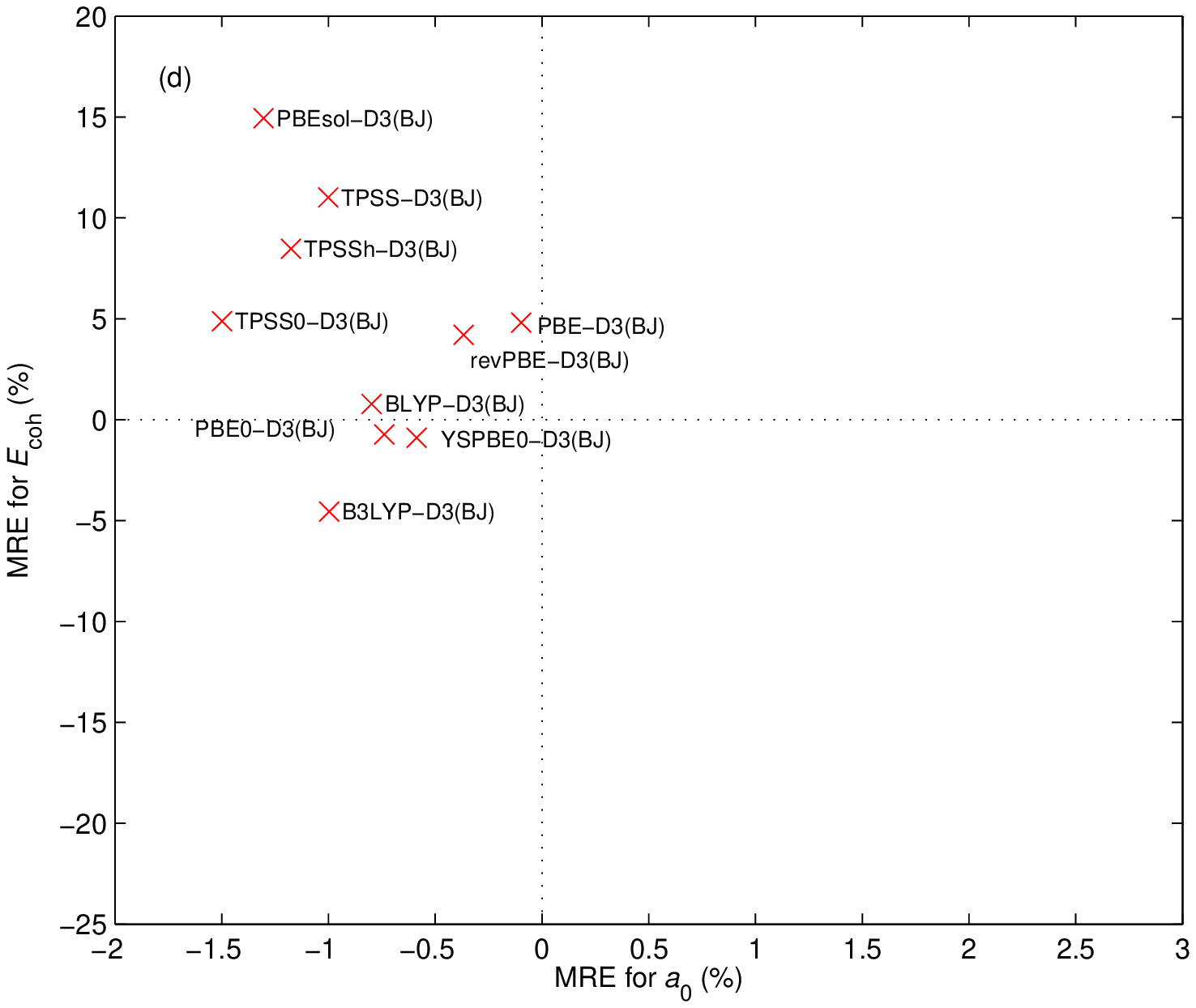}
\caption{\label{fig_mre}MRE of $a_{0}$ versus MRE of $E_{\text{coh}}$
for the (a) LDA and GGA, (b) MGGA, (c) hybrid, and (d) DFT-D3(BJ) functionals.}
\end{figure*}

\begin{figure*}
\includegraphics[scale=0.56]{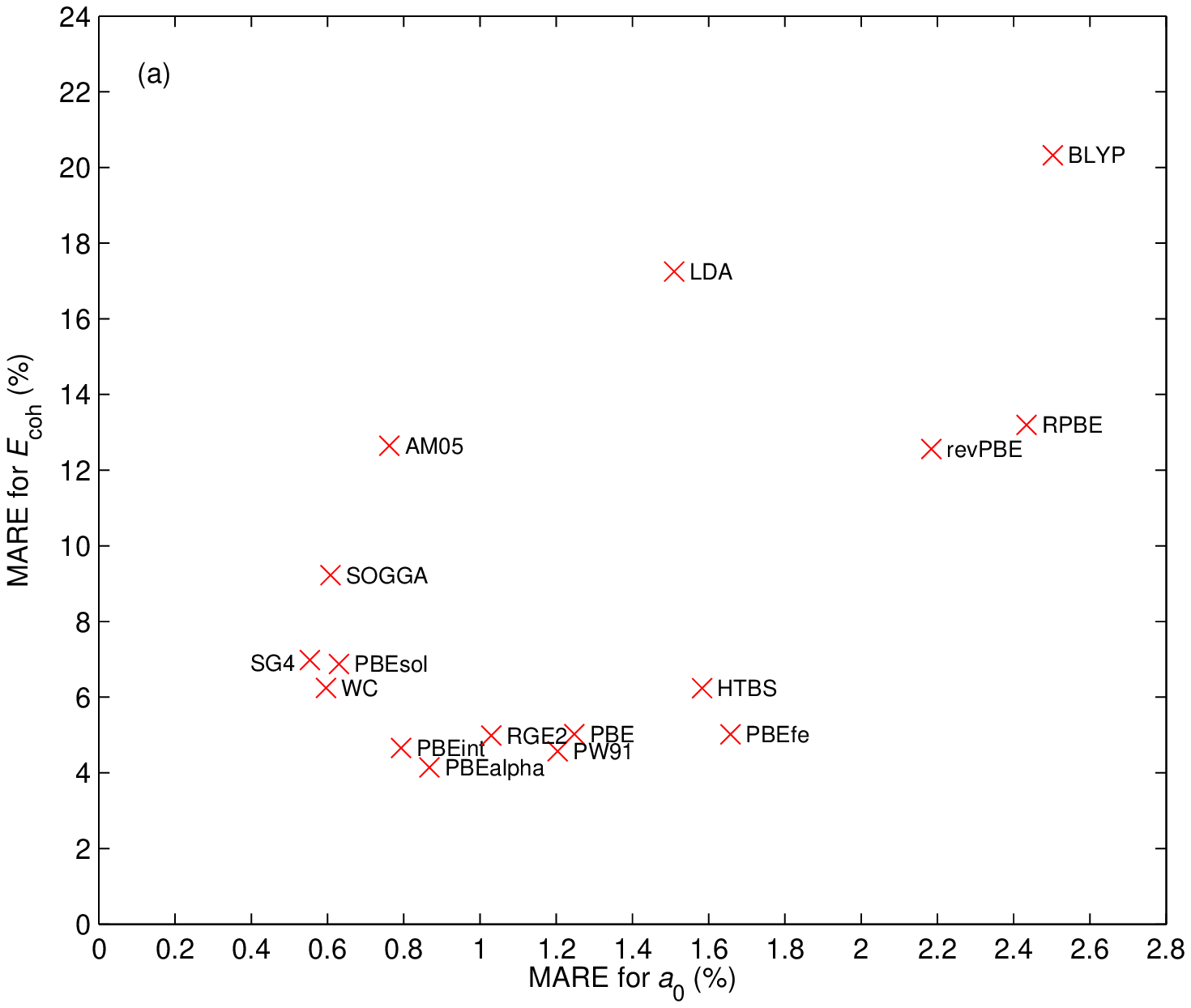}
\includegraphics[scale=0.56]{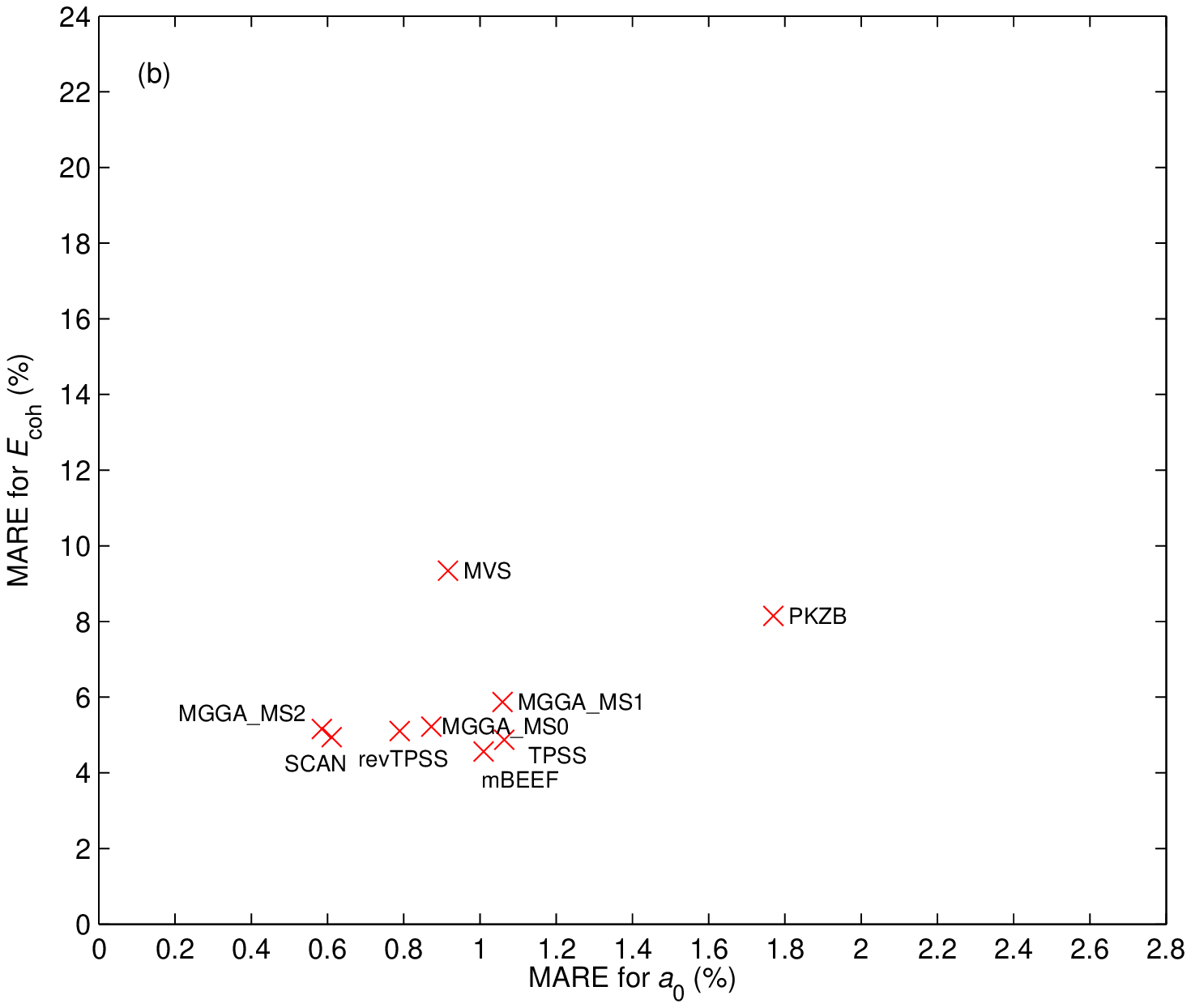}
\includegraphics[scale=0.56]{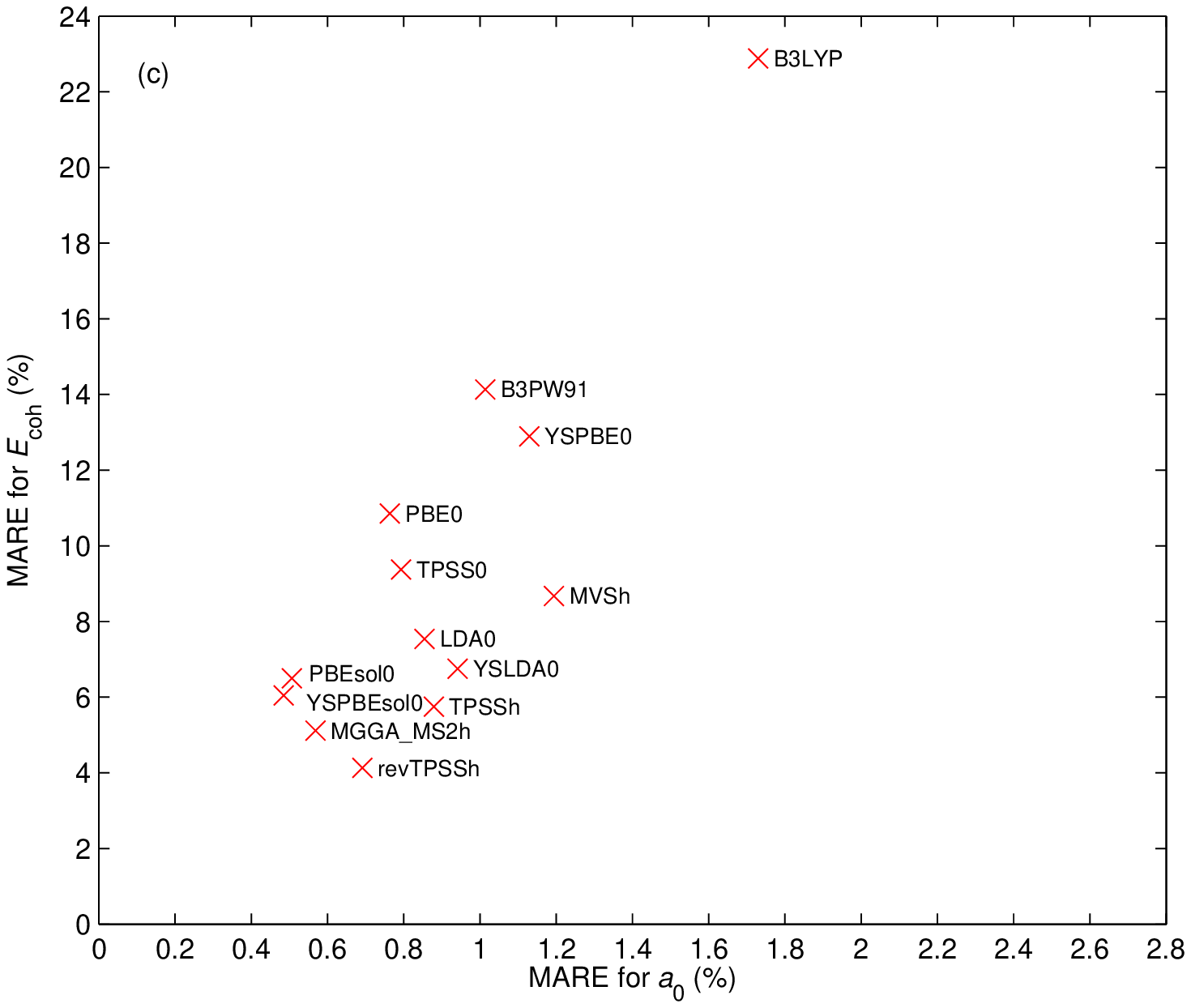}
\includegraphics[scale=0.56]{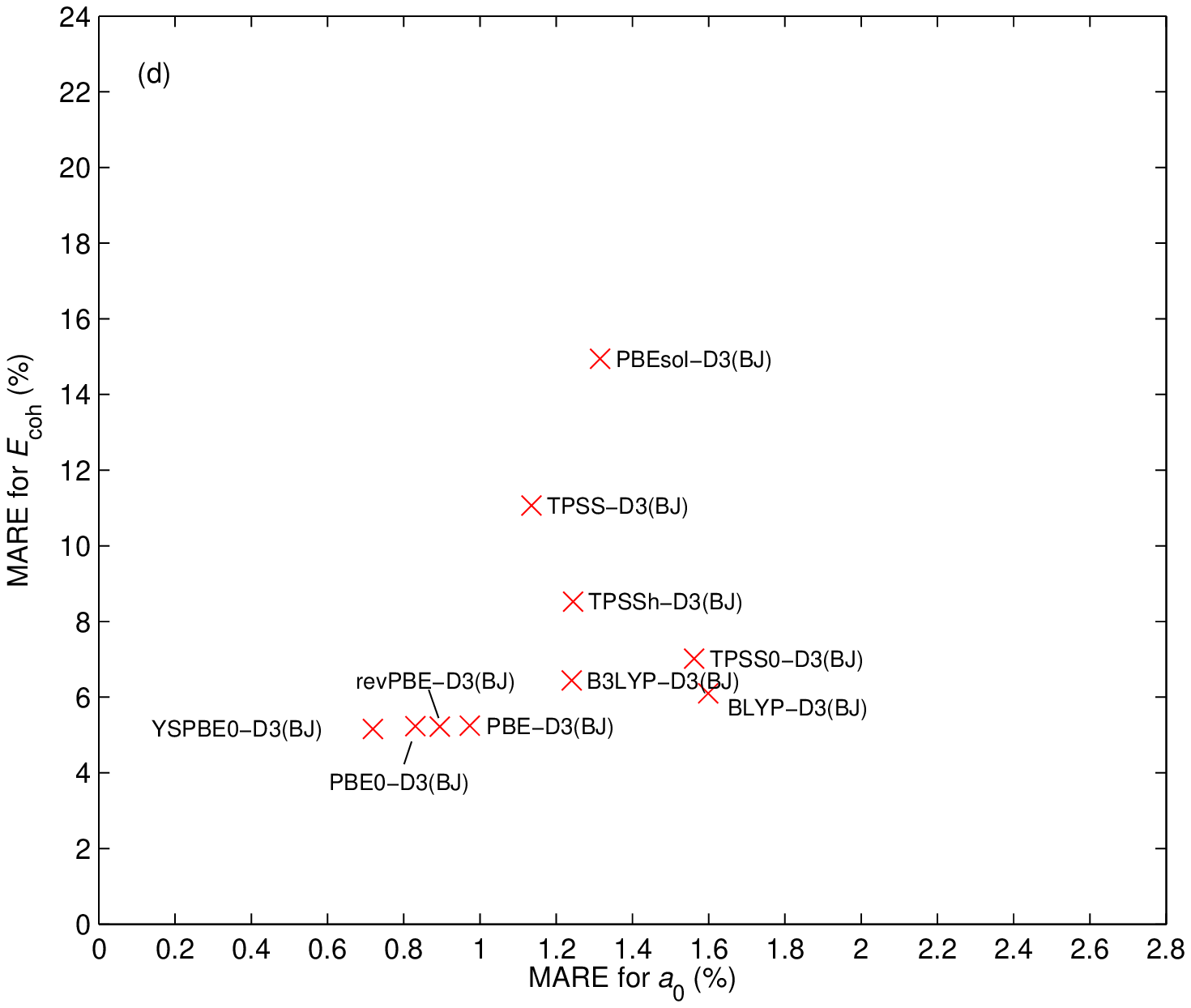}
\caption{\label{fig_mare}MARE of $a_{0}$ versus MARE of $E_{\text{coh}}$
for the (a) LDA and GGA, (b) MGGA, (c) hybrid, and (d) DFT-D3(BJ) functionals.}
\end{figure*}

\begin{figure}
\includegraphics[scale=0.56]{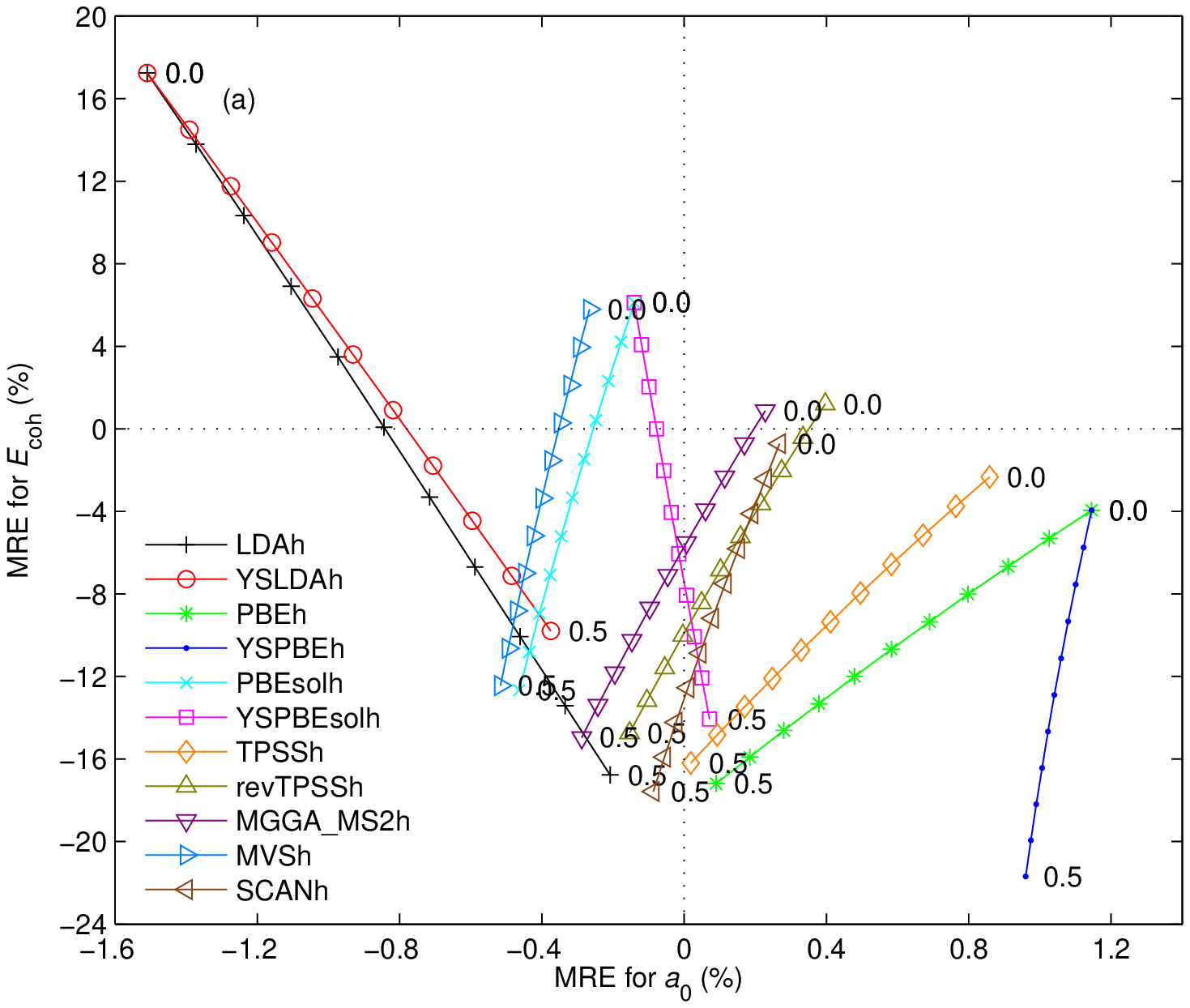}
\includegraphics[scale=0.56]{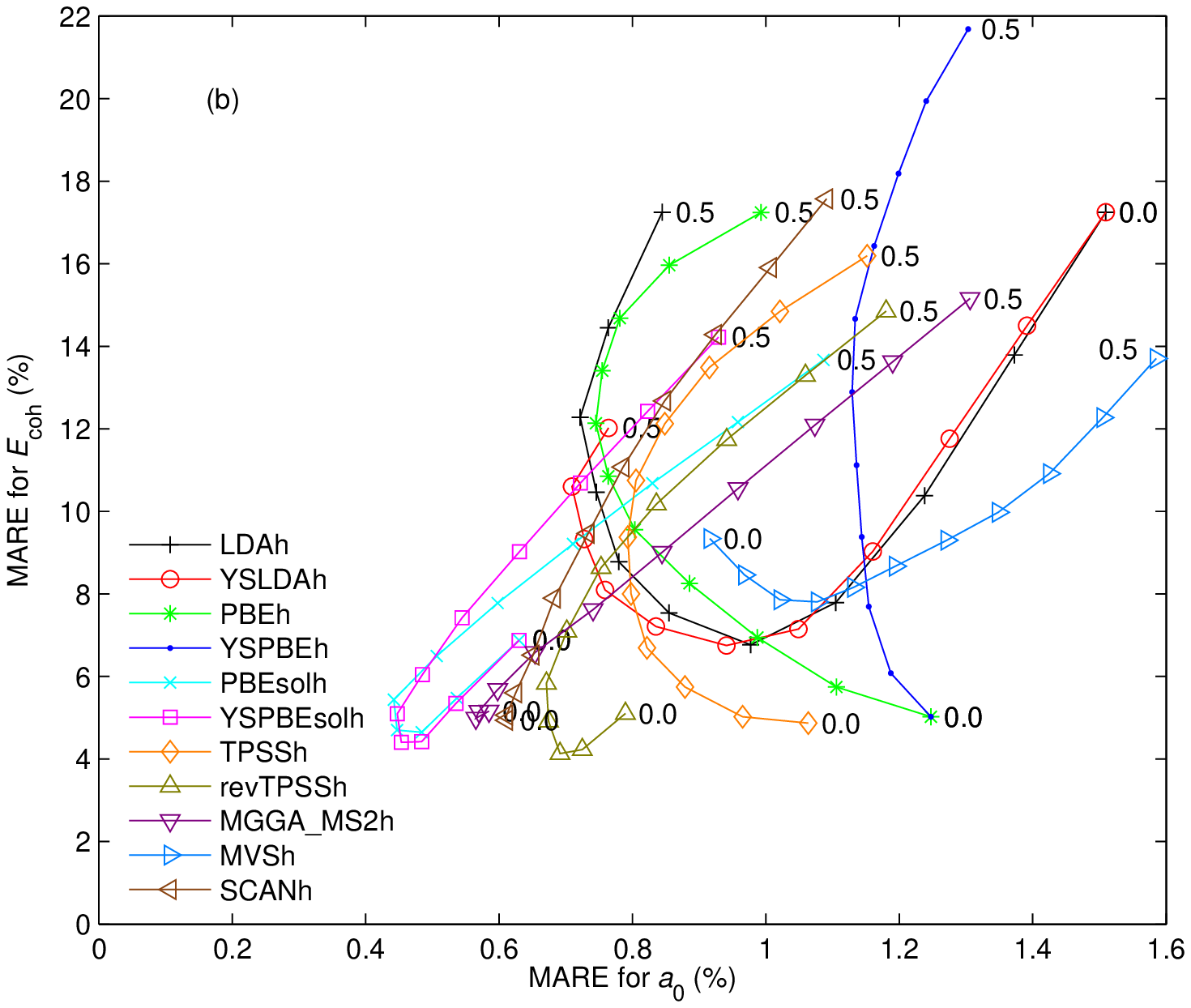}
\caption{\label{fig_mre_mare_group}(a) MRE of $a_{0}$ versus MRE of
$E_{\text{coh}}$ and (b) MARE of $a_{0}$ versus MARE of $E_{\text{coh}}$ for
hybrid functionals with the fraction of HF exchange $\alpha_{\text{x}}$ varied
between 0 and 0.5 with steps of 0.05. The lines connecting the data points are
guide to the eyes.}
\end{figure}

Table~\ref{results_strong} shows the mean error (ME), mean absolute error
(MAE), mean relative error (MRE), and mean absolute error (MARE) on the
equilibrium lattice constant $a_{0}$, bulk modulus $B_{0}$, and cohesive energy
$E_{\text{coh}}$ for the 44 strongly bound solids. Most of the results are also
shown graphically in Figs.~\ref{fig_mre}, and \ref{fig_mare},
which provide a convenient way to compare the
performance of the functionals. The values of $a_{0}$,
$B_{0}$, and $E_{\text{coh}}$ for all solids and functionals can be found in
the supplementary material (SM).\cite{SM_functionals_test}
Since the trends in the MRE/MARE
are similar as for the ME/MAE, the discussion of the results
will be based mainly on the ME and MAE.

We start with the results for the lattice constant and bulk modulus. These two
properties are quite often described with the same accuracy by a functional, but
with opposite trends (i.e., an underestimation of $a_{0}$ is accompanied by an
overestimation of $B_{0}$ or vice-versa), as seen in Figs.~S1-S62 of the SM
where the curves for the relative error for $a_{0}$ (left panel) and $B_{0}$
(middle panel) are approximately like mirror images. The smallest MAE for
$a_{0}$, 0.021~\AA, is obtained by the hybrid-GGA functionals YSPBEsol0 and
PBEsol0, which is in line with the conclusion of Ref.~\onlinecite{SchimkaJCP11}
that combining PBEsol with 25\% of HF exchange improves over PBEsol (one of
the most accurate GGA functionals for this quantity), PBE, and HSE06
($\approx$YSPBE0). YSPBEsol0 performs rather well also for $B_{0}$ with a MAE
of 8.7~GPa, but is not the best method since a couple of other functionals
lead to a MAE around 7.5~GPa, like for example
WC, MGGA\_MS2, SCAN, PBE0, and PBE-D3(BJ).
Note that four functionals (PBEsol, MGGA\_MS2, SCAN, and YSPBEsol0)
lead to a MARE for $B_{0}$ below 7\%.
The functionals which perform very well for both $a_{0}$
(MAE not larger than $\sim0.03$~\AA) and $B_{0}$ (MAE below 9~GPa)
are the GGAs WC, SOGGA, PBEsol, and SG4, the MGGAs
MGGA\_MS2 and SCAN, and the hybrids YSPBEsol0 and MGGA\_MS2h.

Turning now to the results for the cohesive energy $E_{\text{coh}}$, we can
see that the MAE is below $\sim0.2$~eV/atom for a dozen of functionals, e.g.,
the GGAs PW91, PBE, and PBEalpha, the MGGA SCAN, the
hybrid-MGGA revTPSSh, and a few DFT-D3/D3(BJ) methods.
The MAE obtained with YSPBEsol0 and PBEsol0 (the best for the lattice constant)
are slightly larger ($\sim0.27$~eV/atom).

Overall, by considering the
results for the three properties ($a_{0}$, $B_{0}$, and $E_{\text{coh}}$),
the recent MGGAs MGGA\_MS2 and SCAN seem to be the most accurate functionals.
They are among the very best functionals for $B_{0}$ and $E_{\text{coh}}$,
and only YSPBEsol0, PBEsol0, and SG4 are more accurate for $a_{0}$.
Other functionals which are also consistently good for the three properties are the
GGAs WC, PBEsol, PBEalpha, PBEint, and SG4, the hybrids
YSPBEsol0, MGGA\_MS2h, and revTPSSh, and the dispersion-corrected
PBE-D3 and PBE-D3(BJ).

It does not seem to be always necessary to use a functional with an
atom-pairwise dispersion term [D3 or D3(BJ)] for the strongly bound solids.
Actually, adding a dispersion term does not systematically improve the results
(we remind that adding a dispersion term should, in principle, shorten
bond lengths since the London dispersion interactions are attractive).
This is for instance the case with
TPSS, TPSSh, and TPSS0, for which the addition of
D3(BJ) strongly overcorrects the overestimation of $a_{0}$, leading
to large negative ME (and large positive ME for $B_{0}$).
In the case of PBEsol (very small ME for $a_{0}$ and $B_{0}$), adding D3 or
D3(BJ) can only deteriorate the results
since this functional alone does not overestimate the lattice constant on average.
However, a clear improvement is obtained with PBE, revPBE, and BLYP.
We note that none of these dispersion-corrected methods lead, for instance,
to MAE below 0.040~\AA~for $a_{0}$ and 8~GPa for $B_{0}$
at the same time. Furthermore, the MAE for
$B_{0}$ is rather large (above 10~GPa) for many of the dispersion corrected
functionals, including PBE0-D3, which leads to a large MAE of
11.7~GPa despite its MAE for $a_{0}$ is only 0.027~\AA.

Regarding the hybrid functionals, it is instructive to look at how the MRE and
MARE for $a_{0}$ and $E_{\text{coh}}$ vary as functions of the fraction
$\alpha_{\text{x}}$ of HF exchange in Eq.~(\ref{Exchybrid}). This is shown in
Fig.~\ref{fig_mre_mare_group} for most screened and unscreened hybrid
functionals without D3 term, where $\alpha_{\text{x}}$ is varied
between 0 and 0.5 with steps of 0.05.
The trends observed in Fig.~\ref{fig_mre_mare_group}(a) for the MRE
show two different behaviors. For the LDA-based and YSPBEsolh functionals,
the value of the MRE for $a_{0}$ goes in the direction of the positive values
when $\alpha_{\text{x}}$ is increased, while the opposite is observed with the
other functionals. Interestingly, in most cases except PBEsolh and MVSh,
adding a fraction of HF exchange reduces the magnitude of the MRE with respect
to the case $\alpha_{\text{x}}=0$.
For the MARE [Fig.~\ref{fig_mre_mare_group}(b)] the main
observations are the following:
the smallest MARE for $a_{0}$ and $E_{\text{coh}}$ are obtained simultaneously
with more or
less the same value of $\alpha_{\text{x}}$ in the case of
PBEsolh, YSPBEsolh, revTPSSh, MGGA\_MS2h, and SCANh. This optimal 
$\alpha_{\text{x}}$ is $\sim0$ for MGGA\_MS2h and SCANh and $\sim0.15$
for the others. For MGGA\_MS2h and SCANh, it can be argued that
since MGGA functionals are \textit{more nonlocal} than LDA/GGA
(in the sense that the KE density $t$ is probably a truly nonlocal
functional of $\rho$), then less HF exchange is required when combined
with a MGGA. For all other functionals except MVSh, the optimal
$\alpha_{\text{x}}$ is larger for $a_{0}$ than for $E_{\text{coh}}$.
The exception observed with MVSh should be related to the
behavior of its enhancement factor $F_{\text{xc}}$, which has by far the most
negative slope as function of $s$ and $\alpha$, as noticed
above in Fig.~\ref{fig_Fxc2}.

A few words about the functionals that were not considered in this work should
also be added, and in particular about the so-called nonlocal van der Waals
(vdW) functionals,\cite{DionPRL04} which include a term of the form given by
Eq.~(\ref{Ecdisp2}). The first of these functionals which were shown to be,
at least, as accurate as PBE for strongly bound solids, namely optPBE-vdW,
optB88-vdW, and optB86b-vdW, were proposed in
Refs. \onlinecite{KlimesJPCM10,KlimesPRB11}.
In Refs. \onlinecite{KlimesPRB11,SchimkaPRB13,ShulenburgerPRB13,ParkCAP15} it
was shown that compared to PBE, optB88-vdW and optPBE-vdW are slightly better
for the cohesive energy, while optB86b-vdW is slightly better for the lattice
constant. In order to make the SCAN functional more accurate for the treatment
of weak interactions, Peng \textit{et al}.\cite{Peng15} proposed to
add a refitted version of the nonlocal
vdW functional rVV10.\cite{VydrovJCP10,SabatiniPRB13} For a test set of
50 solids, SCAN+rVV10 was shown to perform similarly as SCAN for the lattice
constant, but to increase by about 1\% the MARE for the cohesive energy.

\begin{figure}
\includegraphics[scale=0.39]{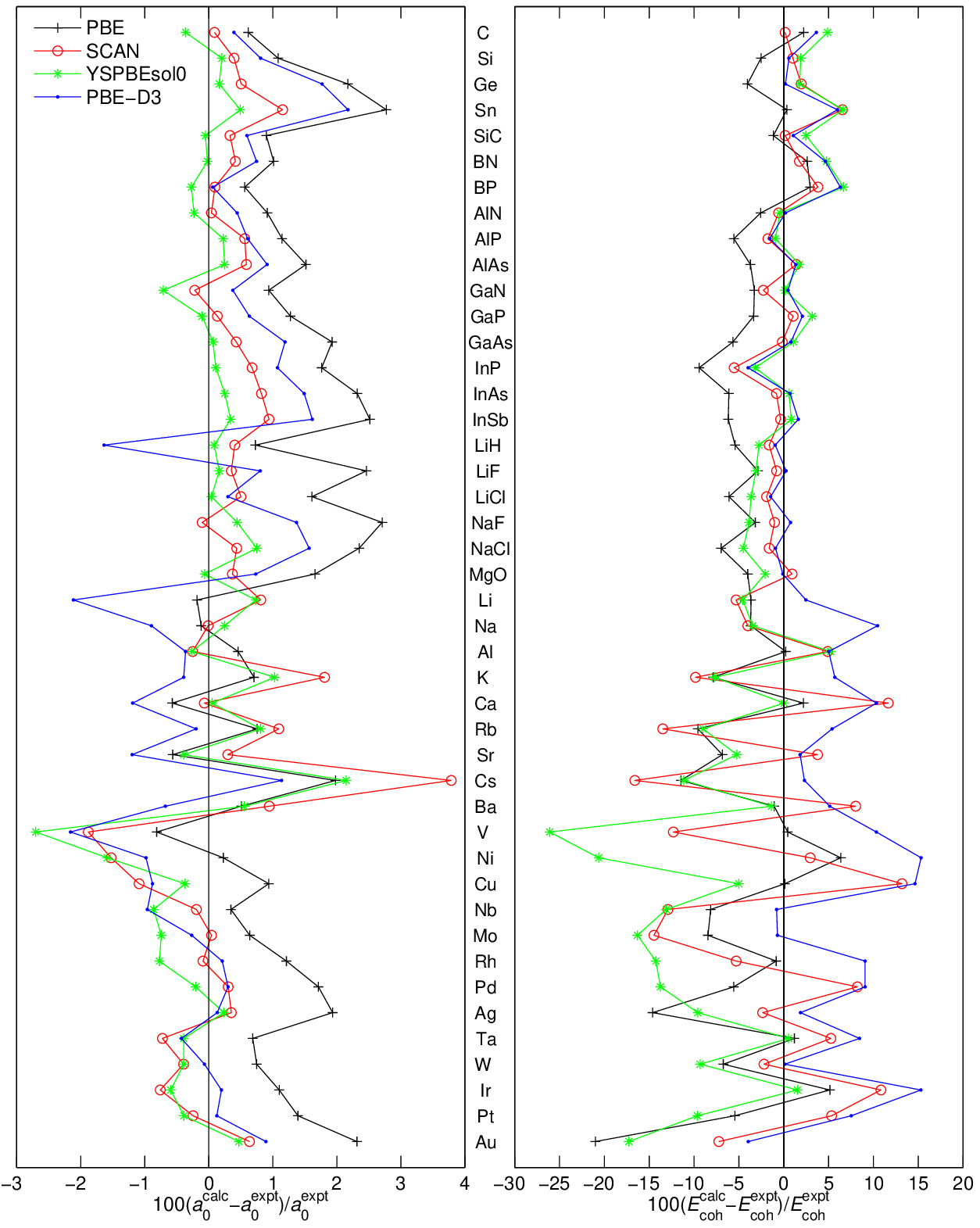}
\caption{\label{fig_solids_1}Relative error (in \%) in the calculated lattice
constant (left panel) and cohesive energy (right panel) for the 44 strongly
bound solids.}
\end{figure}

The detailed results for every solid and functional are shown in the
SM,\cite{SM_functionals_test} and Fig.~\ref{fig_solids_1} gathers the
results for some of the most accurate functionals compared to the
standard PBE. In order to avoid a lengthy discussion, only the
most interesting observations are now discussed. By looking at Figs.~S1-S62,
which show the MRE (in \%) for the lattice constant, bulk modulus, and cohesive
energy, we can immediately see that for many functionals, some of the largest
MRE for $a_{0}$ and $B_{0}$ are found for alkali metals (K, Rb, and Cs),
alkali-earth metals (Ca, Sr, and Ba), and the transition metal V.
For these solids, the MRE for $a_{0}$ increases with the nuclear charge
and can reach 4\%-8\% for Cs. Such large MRE for $a_{0}$ are negative for LDA
(accompanied by an overbinding) and positive (underbinding) for several GGAs
like revPBE or HTBS and, quite interestingly, all (hybrid)-MGGAs.
Such very large overestimations for the heavy alkali metals with TPSS and
revTPSS were already reported.
\cite{CsonkaPRB09,PerdewPRL09,TaoPRB10,SchimkaPRB13}
As argued in Ref.~\onlinecite{TaoPRB10}, the alkali metals are very soft
($B_{0}$ is below 5~GPa) and have a large polarizable core, such that the
long-range core-core dispersion interactions, missing in non-dispersion
corrected functionals, should have a non-negligible effect on the results.
Therefore, it is maybe for the right reason that a semilocal/hybrid
functional, in particular if it is constructed from first-priciples, underbinds
the alkali metals. Adding a D3/D3(BJ) term reduces the error for the alkali
metals, but overcorrects strongly in some cases [e.g., BLYP-D3(BJ) in Fig.~S47].
The results with the DFT-D3/D3(BJ) methods could easily be improved by
tuning the coefficients $C_{n}^{AB}$ in Eq.~(\ref{Ecdisp1}). In the case of
PBE-D3, for instance, it would be possible to strongly reduce the errors
involving the Li atoms by using smaller value for the coefficients, while
for all systems with the diamond or zincblende structures, larger coefficients
would be required.
Such underbinding with the semilocal/hybrid functionals is not observed in the
case of the similar alkali-earth metals, which should be due to the following
reasons: they are slightly less van der Waals like ($B_{0}$ is above 10~GPa)
and the additional valence $s$-electron should reduce the inhomogeneity
in $\rho$, making the semilocal functionals more appropriate. The ionic solids
Li$X$ and Na$X$ are systems for which the MRE can also very large.

Looking at the trends for the $3d$, $4d$, and $5d$ transition metals, most
functionals show the same behavior for the lattice constant; from left to right
within a row (e.g., from Nb to Ag), the MRE goes in the direction of the
positive values.\cite{RopoPRB08,SchimkaPRB13,JanthonJCTC14} This behavior is
the most pronounced for the strong GGAs like revPBE or BLYP [also if D3/D3(BJ)
is included], while it can be strongly reduced with some of the MGGA and hybrid
functionals, similarly as RPA does.\cite{SchimkaPRB13}

A summary of this section on the strongly bound solids is the following.
Among the tested functionals, about 12 of them are in the group of the best
performing for all three properties ($a_{0}$, $B_{0}$, and $E_{\text{coh}}$) at the
same time. This includes GGAs (WC, PBEsol, PBEalpha, PBEint, and SG4), MGGAs
(MGGA\_MS2 and SCAN), hybrids (YSPBEsol0, MGGA\_MS2h, and revTPSSh), and
dispersion-corrected methods [PBE-D3/D3(BJ)]. Therefore, as also shown
more clearly in Figs.~\ref{fig_mre} and \ref{fig_mare},
for every type of approximations except LDA, there are a few functionals
belonging to the group of the best ones. Furthermore, we have also noticed that
MGGA\_MS2 and SCAN give the best results when they are not mixed with HF
exchange, which is a very interesting property from the practical point of
view since the calculation of the HF exchange is very expensive for solids.

\subsection{\label{dispersion}Weakly bound solids}

\begin{figure}
\includegraphics[scale=0.57]{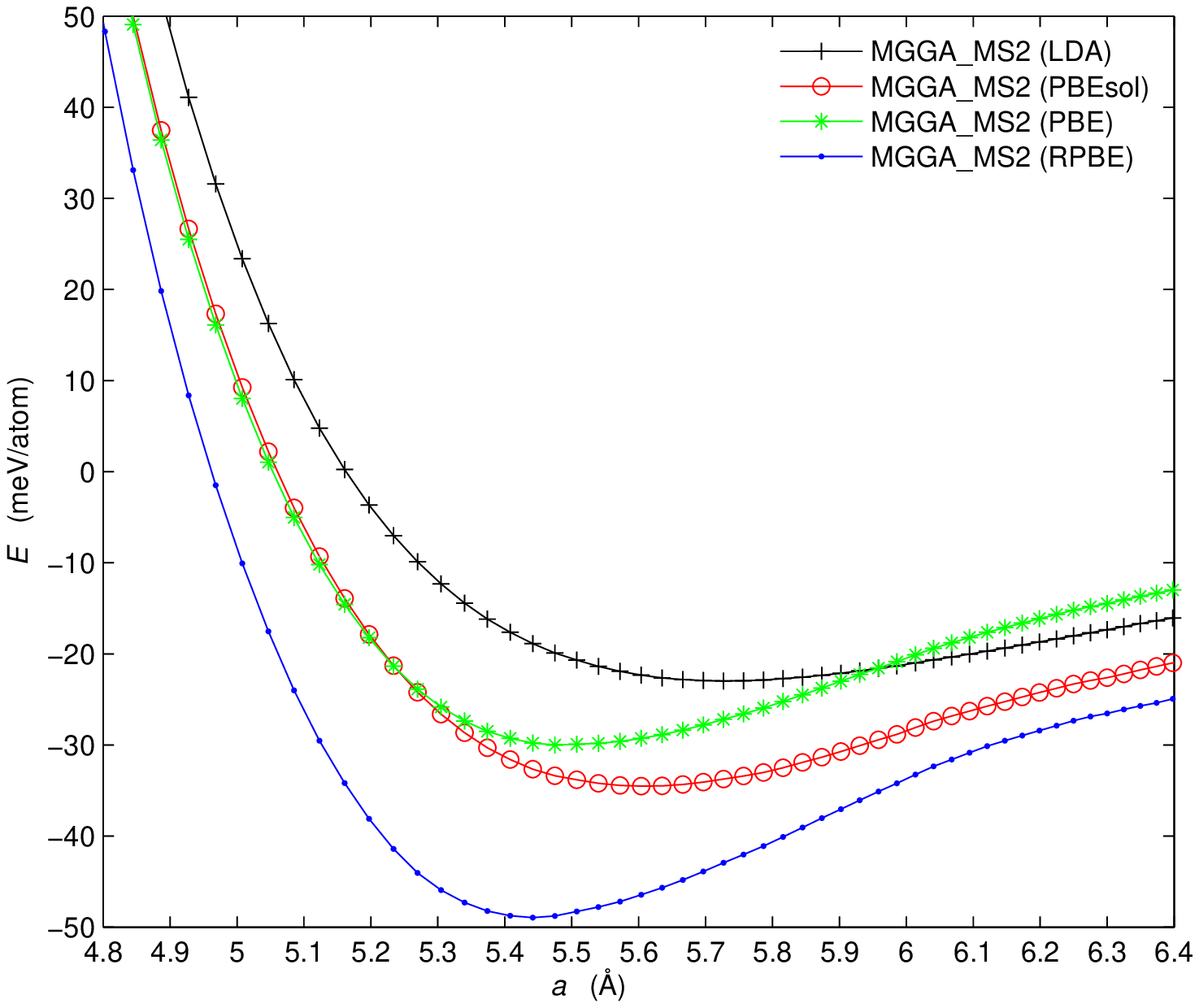}
\caption{\label{fig_Ar}MGGA\_MS2 total energy of Ar plotted as a function of
the lattice constant $a$. The MGGA\_MS2 total-energy functional was evaluated
with orbitals/densities generated from various potentials (indicated in
parenthesis). The zero of the energy axis was chosen such that the cohesive
energy of Ar is given by the value at the minimum of a curve.
The reference CCSD(T) values for $a_{0}$ and $E_{\text{coh}}$ are
5.25~\AA~and 88~meV/atom, respectively.}
\end{figure}

\begingroup
\squeezetable
\begin{table*}
\caption{\label{results_RG}Equilibrium lattice constant $a_{0}$ (in \AA) and
cohesive energy $E_{\text{coh}}$ (in meV/atom and with opposite sign) of
rare-gas solids calculated from various functionals and compared to reference
[CCSD(T), very close to experiment] as well as other methods.
The results for the LDA, GGA, and GGA+D
functionals were obtained from self-consistent calculations, while the PBE
orbitals/density were used for the other functionals. Within each group,
the functionals are ordered by increasing overall error.
For hybrid functionals, the fraction $\alpha_{\text{x}}$ of HF
exchange is indicated in parenthesis.}
{\tiny
\begin{ruledtabular}
\begin{tabular}{lcccccc}
\multicolumn{1}{l}{} &
\multicolumn{2}{c}{Ne} &
\multicolumn{2}{c}{Ar} &
\multicolumn{2}{c}{Kr} \\
\cline{2-3}\cline{4-5}\cline{6-7}
\multicolumn{1}{l}{Functional} &
\multicolumn{1}{c}{$a_{0}$} &
\multicolumn{1}{c}{$E_{\text{coh}}$} &
\multicolumn{1}{c}{$a_{0}$} &
\multicolumn{1}{c}{$E_{\text{coh}}$} &
\multicolumn{1}{c}{$a_{0}$} &
\multicolumn{1}{c}{$E_{\text{coh}}$} \\
\hline
LDA           \\
LDA \cite{PerdewPRB92a} &  3.86 (-10\%) &     87 (234\%) &  4.94 (-6\%) &    138 (57\%) &  5.33 (-5\%) &    169 (39\%) \\
\hline
GGA           \\
PBEalpha \cite{MadsenPRB07} &  4.39 (2\%) &     23 (-10\%) &  5.59 (6\%) &     33 (-62\%) &  5.97 (7\%) &     43 (-65\%) \\
SOGGA \cite{ZhaoJCP08} &  4.52 (5\%) &     23 (-11\%) &  5.77 (10\%) &     29 (-67\%) &  6.14 (10\%) &     35 (-71\%) \\
RPBE \cite{HammerPRB99} &  4.74 (10\%) &     26 (0\%) &  6.25 (19\%) &     28 (-68\%) &  6.83 (22\%) &     29 (-76\%) \\
PBE \cite{PerdewPRL96} &  4.60 (7\%) &     19 (-26\%) &  5.96 (13\%) &     23 (-73\%) &  6.42 (15\%) &     27 (-78\%) \\
HTBS \cite{HaasPRB11} &  4.80 (12\%) &     23 (-12\%) &  6.34 (21\%) &     25 (-72\%) &  6.93 (24\%) &     26 (-79\%) \\
PW91 \cite{PerdewPRB92b} &  4.62 (7\%) &     47 (82\%) &  6.05 (15\%) &     49 (-44\%) &  6.55 (17\%) &     51 (-58\%) \\
PBEsol \cite{PerdewPRL08} &  4.70 (9\%) &     12 (-54\%) &  5.88 (12\%) &     17 (-81\%) &  6.13 (10\%) &     23 (-81\%) \\
PBEint \cite{FabianoPRB10} &  4.78 (11\%) &     14 (-46\%) &  6.21 (18\%) &     17 (-81\%) &  6.67 (19\%) &     20 (-84\%) \\
RGE2 \cite{RuzsinszkyJCTC09} &  4.92 (14\%) &     14 (-45\%) &  6.43 (23\%) &     16 (-81\%) &  6.99 (25\%) &     18 (-85\%) \\
WC \cite{WuPRB06} &  4.87 (13\%) &     12 (-54\%) &  6.34 (21\%) &     14 (-84\%) &  6.86 (23\%) &     16 (-87\%) \\
PBEfe \cite{SarmientoPerezJCTC15} &  3.88 (-10\%) &     99 (280\%) &  5.00 (-5\%) &    152 (73\%) &  5.42 (-3\%) &    184 (51\%) \\
SG4 \cite{ConstantinPRB16} &  5.25 (22\%) &      9 (-67\%) &  $ > $ 6.6 ($ > $ 26\%) &       &  $ > $ 7.1 ($ > $ 27\%) &       \\
revPBE \cite{ZhangPRL98} &  5.31 (24\%) &      7 (-74\%) &  $ > $ 6.6 ($ > $ 26\%) &       &  $ > $ 7.1 ($ > $ 27\%) &       \\
BLYP \cite{BeckePRA88,LeePRB88} &  $ > $ 5.6 ($ > $ 31\%) &       &  $ > $ 6.6 ($ > $ 26\%) &       &  $ > $ 7.1 ($ > $ 27\%) &       \\
AM05 \cite{ArmientoPRB05} &  $ > $ 5.6 ($ > $ 31\%) &       &  $ > $ 6.6 ($ > $ 26\%) &       &  $ > $ 7.1 ($ > $ 27\%) &       \\
\hline
MGGA          \\
MGGA\_MS2 \cite{SunJCP13} &  4.31 (0\%) &     26 (1\%) &  5.48 (4\%) &     30 (-66\%) &  5.96 (6\%) &     45 (-63\%) \\
MGGA\_MS1 \cite{SunJCP13} &  4.34 (1\%) &     26 (0\%) &  5.58 (6\%) &     27 (-69\%) &  6.10 (9\%) &     40 (-67\%) \\
MGGA\_MS0 \cite{SunJCP12} &  4.16 (-3\%) &     40 (53\%) &  5.41 (3\%) &     46 (-47\%) &  5.89 (5\%) &     61 (-50\%) \\
SCAN \cite{SunPRL15} &  4.03 (-6\%) &     54 (107\%) &  5.31 (1\%) &     61 (-30\%) &  5.74 (2\%) &     72 (-41\%) \\
PKZB \cite{PerdewPRL99} &  4.66 (9\%) &     27 (2\%) &  6.20 (18\%) &     26 (-70\%) &  6.76 (21\%) &     30 (-75\%) \\
MVS \cite{SunPNAS15} &  4.02 (-6\%) &     59 (125\%) &  5.41 (3\%) &     56 (-37\%) &  5.79 (3\%) &     69 (-43\%) \\
TPSS \cite{TaoPRL03} &  4.92 (15\%) &     11 (-59\%) &  6.45 (23\%) &     11 (-87\%) &  6.98 (25\%) &     15 (-88\%) \\
mBEEF \cite{WellendorffJCP14} &  3.92 (-9\%) &    134 (416\%) &  5.26 (0\%) &    142 (62\%) &  5.75 (3\%) &    161 (32\%) \\
revTPSS \cite{PerdewPRL09} &  5.05 (17\%) &      7 (-72\%) &  $ > $ 6.6 ($ > $ 26\%) &       &  7.04 (26\%) &     12 (-90\%) \\
\hline
hybrid-LDA    \\
LDA0 \cite{PerdewPRB92a,PerdewJCP96} (0.25) &  4.00 (-7\%) &     51 (96\%) &  5.18 (-1\%) &     71 (-19\%) &  5.57 (0\%) &     90 (-26\%) \\
YSLDA0 \cite{PerdewPRB92a,PerdewJCP96} (0.25) &  3.96 (-8\%) &     60 (131\%) &  5.12 (-2\%) &     86 (-2\%) &  5.51 (-1\%) &    108 (-11\%) \\
\hline
hybrid-GGA    \\
PBE0 \cite{ErnzerhofJCP99,AdamoJCP99} (0.25) &  4.61 (7\%) &     11 (-57\%) &  5.96 (14\%) &     15 (-83\%) &  6.41 (15\%) &     19 (-84\%) \\
PBEsol0 \cite{PerdewPRL08,PerdewJCP96} (0.25) &  4.66 (8\%) &      7 (-75\%) &  5.79 (10\%) &     12 (-86\%) &  6.06 (8\%) &     20 (-84\%) \\
YSPBE0 \cite{KrukauJCP06,TranPRB11} (0.25) &  4.76 (11\%) &     10 (-60\%) &  6.21 (18\%) &     14 (-84\%) &  6.69 (19\%) &     17 (-86\%) \\
YSPBEsol0 \cite{SchimkaJCP11} (0.25) &  4.93 (15\%) &      5 (-81\%) &  6.26 (19\%) &      8 (-91\%) &  6.55 (17\%) &     11 (-91\%) \\
B3LYP \cite{StephensJPC94} (0.20) &  $ > $ 5.6 ($ > $ 31\%) &       &  $ > $ 6.6 ($ > $ 26\%) &       &  $ > $ 7.1 ($ > $ 27\%) &       \\
B3PW91 \cite{BeckeJCP93} (0.20) &  $ > $ 5.6 ($ > $ 31\%) &       &  $ > $ 6.6 ($ > $ 26\%) &       &  $ > $ 7.1 ($ > $ 27\%) &       \\
\hline
hybrid-MGGA   \\
MGGA\_MS2h \cite{SunJCP13} (0.09) &  4.31 (0\%) &     23 (-11\%) &  5.48 (4\%) &     27 (-70\%) &  5.97 (7\%) &     41 (-67\%) \\
MVSh \cite{SunPNAS15} (0.25) &  4.05 (-6\%) &     40 (55\%) &  5.44 (4\%) &     40 (-55\%) &  5.83 (4\%) &     52 (-58\%) \\
TPSSh \cite{StaroverovJCP03} (0.10) &  4.93 (15\%) &      8 (-67\%) &  6.46 (23\%) &      9 (-90\%) &  6.98 (25\%) &     13 (-90\%) \\
TPSS0 \cite{TaoPRL03,PerdewJCP96} (0.25) &  4.96 (15\%) &      5 (-80\%) &  6.47 (23\%) &      6 (-93\%) &  6.98 (25\%) &      9 (-92\%) \\
revTPSSh \cite{CsonkaJCTC10} (0.10) &  5.06 (18\%) &      5 (-79\%) &  $ > $ 6.6 ($ > $ 26\%) &       &  7.03 (26\%) &     10 (-92\%) \\
\hline
GGA+D         \\
revPBE-D3(BJ) \cite{GrimmeJCC11} &  4.80 (12\%) &     25 (-2\%) &  5.67 (8\%) &     82 (-7\%) &  5.96 (7\%) &    126 (3\%) \\
PBEsol-D3(BJ) \cite{GoerigkPCCP11} &  4.59 (7\%) &     22 (-16\%) &  5.46 (4\%) &     71 (-19\%) &  5.69 (2\%) &    116 (-5\%) \\
revPBE-D3 \cite{GrimmeJCP10} &  4.73 (10\%) &     25 (-4\%) &  5.64 (7\%) &     68 (-23\%) &  5.85 (5\%) &    109 (-11\%) \\
PBE-D3(BJ) \cite{GrimmeJCC11} &  4.46 (4\%) &     37 (42\%) &  5.49 (5\%) &     86 (-2\%) &  5.85 (5\%) &    117 (-4\%) \\
PBEsol-D3 \cite{GoerigkPCCP11} &  4.53 (5\%) &     29 (13\%) &  5.37 (2\%) &     61 (-31\%) &  5.58 (0\%) &    102 (-17\%) \\
BLYP-D3 \cite{GrimmeJCP10} &  4.25 (-1\%) &     16 (-38\%) &  5.35 (2\%) &     70 (-21\%) &  5.70 (2\%) &    127 (4\%) \\
PBE-D3 \cite{GrimmeJCP10} &  4.39 (2\%) &     46 (78\%) &  5.58 (6\%) &     83 (-6\%) &  5.90 (5\%) &    113 (-8\%) \\
BLYP-D3(BJ) \cite{GrimmeJCC11} &  4.58 (7\%) &      3 (-89\%) &  5.37 (2\%) &     71 (-19\%) &  5.67 (1\%) &    134 (10\%) \\
RPBE-D3 \cite{DFTD3} &  4.49 (4\%) &     52 (101\%) &  5.66 (8\%) &     91 (3\%) &  6.03 (8\%) &    116 (-5\%) \\
\hline
MGGA+D        \\
TPSS-D3(BJ) \cite{GrimmeJCC11} &  4.69 (9\%) &     28 (7\%) &  5.67 (8\%) &     78 (-11\%) &  5.97 (7\%) &    118 (-3\%) \\
TPSS-D3 \cite{GrimmeJCP10} &  4.53 (6\%) &     36 (39\%) &  5.69 (8\%) &     76 (-14\%) &  5.99 (7\%) &    111 (-9\%) \\
MGGA\_MS2-D3 \cite{SunJCP13} &  4.19 (-3\%) &     55 (113\%) &  5.43 (3\%) &     83 (-6\%) &  5.91 (5\%) &    105 (-14\%) \\
MGGA\_MS1-D3 \cite{SunJCP13} &  4.14 (-4\%) &     60 (132\%) &  5.37 (2\%) &     96 (9\%) &  5.83 (4\%) &    121 (-1\%) \\
MGGA\_MS0-D3 \cite{SunJCP13} &  4.10 (-5\%) &     70 (170\%) &  5.38 (2\%) &     98 (11\%) &  5.85 (4\%) &    120 (-2\%) \\
\hline
hybrid-GGA+D  \\
PBE0-D3(BJ) \cite{GrimmeJCC11} (0.25) &  4.45 (4\%) &     28 (7\%) &  5.46 (4\%) &     82 (-7\%) &  5.79 (3\%) &    121 (-1\%) \\
B3LYP-D3 \cite{GrimmeJCP10} (0.20) &  4.25 (-1\%) &     23 (-10\%) &  5.30 (1\%) &     68 (-23\%) &  5.61 (0\%) &    129 (5\%) \\
YSPBE0-D3(BJ) \cite{DFTD3} (0.25) &  4.62 (8\%) &     23 (-11\%) &  5.66 (8\%) &     75 (-15\%) &  5.98 (7\%) &    116 (-5\%) \\
PBE0-D3 \cite{GrimmeJCP10} (0.25) &  4.39 (2\%) &     36 (40\%) &  5.45 (4\%) &     74 (-16\%) &  5.71 (2\%) &    114 (-6\%) \\
B3LYP-D3(BJ) \cite{GrimmeJCC11} (0.20) &  4.39 (2\%) &     12 (-52\%) &  5.32 (1\%) &     78 (-12\%) &  5.65 (1\%) &    135 (11\%) \\
YSPBE0-D3 \cite{DFTD3} (0.25) &  4.46 (4\%) &     33 (25\%) &  5.75 (9\%) &     60 (-32\%) &  6.21 (11\%) &     74 (-39\%) \\
\hline
hybrid-MGGA+D \\
TPSSh-D3(BJ) \cite{HoffmannJCC14} (0.10) &  4.69 (9\%) &     25 (-3\%) &  5.65 (8\%) &     79 (-10\%) &  5.94 (6\%) &    122 (0\%) \\
TPSS0-D3 \cite{GrimmeJCP10} (0.25) &  4.53 (5\%) &     29 (13\%) &  5.64 (7\%) &     66 (-25\%) &  5.78 (3\%) &    108 (-12\%) \\
TPSS0-D3(BJ) \cite{GrimmeJCC11} (0.25) &  4.66 (8\%) &     21 (-18\%) &  5.57 (6\%) &     70 (-21\%) &  5.81 (4\%) &    112 (-8\%) \\
TPSSh-D3 \cite{GoerigkPCCP11} (0.10) &  4.55 (6\%) &     33 (28\%) &  5.70 (9\%) &     70 (-20\%) &  5.91 (6\%) &    108 (-12\%) \\
MGGA\_MS2h-D3 \cite{SunJCP13} (0.09) &  4.18 (-3\%) &     52 (100\%) &  5.44 (4\%) &     79 (-10\%) &  5.91 (5\%) &    101 (-17\%) \\
\hline
Previous works \\
optB88-vdW \cite{KlimesJPCM10} (Ref.~\onlinecite{TranJCP13}) &  4.24 (-1\%) &     59 (127\%) &  5.24 (0\%) &    143 (62\%) &  5.63 (1\%) &    181 (48\%) \\
C09x-vdW \cite{CooperPRB10} (Ref.~\onlinecite{TranJCP13}) &  4.50 (5\%) &     62 (138\%) &  5.33 (2\%) &    128 (45\%) &  5.64 (1\%) &    163 (34\%) \\
rVV10 \cite{VydrovJCP10,SabatiniPRB13} (Ref.~\onlinecite{TranJCP13}) &  4.19 (-2\%) &     49 (88\%) &  5.17 (-2\%) &    117 (33\%) &  5.53 (-1\%) &    162 (33\%) \\
rev-vdW-DF2 \cite{HamadaPRB14} (Ref.~\onlinecite{CallsenPRB15}) &  4.43 (3\%) &     30 (15\%) &  5.35 (2\%) &     90 (2\%) &  5.71 (2\%) &    120 (-2\%) \\
PBE+TS \cite{TkatchenkoPRL09} (Ref.~\onlinecite{AlSaidiJCTC12}) &  4.42 (3\%) &     43 (65\%) &  5.51 (5\%) &     83 (-6\%) &  5.90 (5\%) &     97 (-20\%) \\
RPA (Ref. \onlinecite{HarlPRB08}) &   4.5 (5\%) &     17 (-35\%) &   5.3 (1\%) &     83 (-6\%) &   5.7 (2\%) &    112 (-8\%) \\
Expt. (Ref. \onlinecite{RosciszewskiPRB00}) &  4.29 &     26 &  5.25 &     88 &  5.63 &    122 \\
CCSD(T) (Ref. \onlinecite{RosciszewskiPRB00}) &  4.30 &     26 &  5.25 &     88 &  5.60 &    122 \\
\end{tabular}
\end{ruledtabular}
}
\end{table*}
\endgroup

\begingroup
\squeezetable
\begin{table*}
\caption{\label{results_LC}Equilibrium lattice constant $c_{0}$ (in \AA) and
interlayer binding energy $E_{\text{b}}$ (in meV/atom and with opposite sign) of
layered solids calculated from various functionals and compared to reference
(RPA) as well as other methods. The results for the LDA, GGA, and GGA+D
functionals were obtained from self-consistent calculations, while the PBE
orbitals/density were used for the other functionals. Within each group,
the functionals are ordered by increasing overall error.
For hybrid functionals, the fraction $\alpha_{\text{x}}$ of HF
exchange is indicated in parenthesis.}
{\tiny
\begin{ruledtabular}
\begin{tabular}{lcccc}
\multicolumn{1}{l}{} &
\multicolumn{2}{c}{Graphite} &
\multicolumn{2}{c}{h-BN} \\
\cline{2-3}\cline{4-5}
\multicolumn{1}{l}{Functional} &
\multicolumn{1}{c}{$c_{0}$} &
\multicolumn{1}{c}{$E_{\text{b}}$} &
\multicolumn{1}{c}{$c_{0}$} &
\multicolumn{1}{c}{$E_{\text{b}}$} \\
\hline
LDA           \\
LDA \cite{PerdewPRB92a} &   6.7 (0\%) &     24 (-50\%) &   6.5 (-3\%) &     28 (-28\%) \\
\hline
GGA           \\
PBEfe \cite{SarmientoPerezJCTC15} &   7.0 (5\%) &     21 (-57\%) &   6.9 (3\%) &     24 (-39\%) \\
SOGGA \cite{ZhaoJCP08} &   7.3 (9\%) &      4 (-91\%) &   7.0 (5\%) &      7 (-83\%) \\
PBEsol \cite{PerdewPRL08} &   7.3 (9\%) &      4 (-92\%) &   7.0 (6\%) &      6 (-84\%) \\
PBEalpha \cite{MadsenPRB07} &   7.6 (14\%) &      4 (-91\%) &   7.3 (10\%) &      6 (-84\%) \\
PBE \cite{PerdewPRL96} &   $\sim8.8$ (31\%) &      1 (-97\%) &   $\sim8.5$ (28\%) &      2 (-94\%) \\
PW91 \cite{PerdewPRB92b} &   $\sim9.3$ (38\%) &      2 (-95\%) &   $\sim9.0$ (36\%) &      3 (-93\%) \\
PBEint \cite{FabianoPRB10} &   $\sim9.3$ (39\%) &      1 (-98\%) &   $\sim9.0$ (35\%) &      2 (-96\%) \\
WC \cite{WuPRB06} &   $\sim9.7$ (45\%) &      1 (-99\%) &   $\sim9.5$ (42\%) &      1 (-97\%) \\
RPBE \cite{HammerPRB99} &   $\sim9.8$ (46\%) &      1 (-97\%) &   $\sim9.8$ (47\%) &      2 (-96\%) \\
HTBS \cite{HaasPRB11} &   $\sim9.9$ (48\%) &      1 (-98\%) &   $\sim9.9$ (49\%) &      2 (-96\%) \\
RGE2 \cite{RuzsinszkyJCTC09} &  $\sim10.0$ (49\%) &      1 (-99\%) &  $\sim10.0$ (49\%) &      1 (-97\%) \\
SG4 \cite{ConstantinPRB16} &  $\sim11.0$ (64\%) &      0 (-99\%) &  $\sim11.0$ (65\%) &      1 (-98\%) \\
revPBE \cite{ZhangPRL98} &  $\sim11.3$ (69\%) &      0 (-99\%) &  $\sim11.3$ (70\%) &      1 (-98\%) \\
AM05 \cite{ArmientoPRB05} &  $ > $ 19 ($ > $ 176\%) &       & $>$ 19 ($>$ 188\%)  &      \\
BLYP \cite{BeckePRA88,LeePRB88} &  $ > $ 19 ($ > $ 176\%) &       &  $>$ 19 ($>$ 188\%)  &       \\
\hline
MGGA          \\
MVS \cite{SunPNAS15} &   6.6 (-1\%) &     32 (-34\%) &   6.4 (-4\%) &     38 (-4\%) \\
SCAN \cite{SunPRL15} &   6.9 (3\%) &     20 (-59\%) &   6.8 (2\%) &     21 (-46\%) \\
mBEEF \cite{WellendorffJCP14} &   7.8 (16\%) &     13 (-72\%) &   7.7 (16\%) &     14 (-63\%) \\
MGGA\_MS2 \cite{SunJCP13} &   7.2 (7\%) &      8 (-83\%) &   7.0 (5\%) &     10 (-74\%) \\
MGGA\_MS0 \cite{SunJCP12} &   7.4 (11\%) &      8 (-83\%) &   7.3 (9\%) &      9 (-76\%) \\
MGGA\_MS1 \cite{SunJCP13} &   7.8 (16\%) &      5 (-90\%) &   7.7 (16\%) &      6 (-86\%) \\
PKZB \cite{PerdewPRL99} &   7.9 (19\%) &      4 (-91\%) &   7.9 (18\%) &      5 (-88\%) \\
revTPSS \cite{PerdewPRL09} &   $ > $ 19 ($ > $ 176\%) &      &   $\sim9.8$ (47\%) &      1 (-98\%) \\
TPSS \cite{TaoPRL03} &   $ > $ 19 ($ > $ 176\%) &       &   $\sim9.9$ (49\%) &      1 (-98\%) \\
\hline
hybrid-LDA    \\
YSLDA0 \cite{PerdewPRB92a,PerdewJCP96} (0.25) &   7.0 (4\%) &     15 (-70\%) &   6.7 (1\%) &     18 (-53\%) \\
LDA0 \cite{PerdewPRB92a,PerdewJCP96} (0.25) &   7.1 (5\%) &     12 (-74\%) &   6.8 (2\%) &     16 (-60\%) \\
\hline
hybrid-GGA    \\
PBEsol0 \cite{PerdewPRL08,PerdewJCP96} (0.25) &   7.3 (9\%) &      5 (-90\%) &   7.0 (5\%) &      7 (-81\%) \\
YSPBEsol0 \cite{SchimkaJCP11} (0.25) &   $\sim7.8$ (17\%) &      1 (-97\%) &   $\sim7.3$ (10\%) &      3 (-92\%) \\
PBE0 \cite{ErnzerhofJCP99,AdamoJCP99} (0.25) &   $\sim8.4$ (25\%) &      2 (-97\%) &   $\sim8.0$ (20\%) &      3 (-94\%) \\
YSPBE0 \cite{KrukauJCP06,TranPRB11} (0.25) &   $\sim9.3$ (39\%) &      1 (-98\%) &   $\sim9.0$ (36\%) &      1 (-97\%) \\
B3LYP \cite{StephensJPC94} (0.20) &  $ > $ 19 ($ > $ 176\%) &       &  $>$ 19 ($>$ 188\%) &      \\
B3PW91 \cite{BeckeJCP93} (0.20) &  $ > $ 19 ($ > $ 176\%) &       & $>$ 19 ($>$ 188\%)  &      \\
\hline
hybrid-MGGA   \\
MVSh \cite{SunPNAS15} (0.25) &   6.7 (0\%) &     25 (-48\%) &   6.5 (-3\%) &     31 (-22\%) \\
MGGA\_MS2h \cite{SunJCP13} (0.09) &   7.2 (7\%) &      8 (-83\%) &   7.0 (5\%) &     10 (-74\%) \\
revTPSSh \cite{CsonkaJCTC10} (0.10) &   $>$ 19 ($>$ 176\%) &      &   $\sim9.4$ (41\%) &      1 (-98\%) \\
TPSSh \cite{StaroverovJCP03} (0.10) &   $>$ 19 ($>$ 176\%) &      &   $\sim9.8$ (48\%) &      1 (-98\%) \\
TPSS0 \cite{TaoPRL03,PerdewJCP96} (0.25) &  $ > $ 19 ($ > $ 176\%) &       &   $\sim9.6$ (44\%) &      1 (-98\%) \\
\hline
GGA+D         \\
RPBE-D3 \cite{DFTD3} &   6.8 (1\%) &     39 (-19\%) &   6.7 (0\%) &     39 (0\%) \\
PBE-D3(BJ) \cite{GrimmeJCC11} &   6.8 (2\%) &     43 (-10\%) &   6.7 (0\%) &     44 (12\%) \\
PBEsol-D3 \cite{GoerigkPCCP11} &   6.7 (0\%) &     38 (-20\%) &   6.6 (-2\%) &     42 (8\%) \\
PBE-D3 \cite{GrimmeJCP10} &   7.1 (5\%) &     39 (-19\%) &   6.8 (3\%) &     41 (5\%) \\
PBEsol-D3(BJ) \cite{GoerigkPCCP11} &   6.7 (-1\%) &     52 (8\%) &   6.5 (-3\%) &     53 (36\%) \\
revPBE-D3 \cite{GrimmeJCP10} &   6.6 (-2\%) &     53 (10\%) &   6.5 (-2\%) &     52 (33\%) \\
BLYP-D3 \cite{GrimmeJCP10} &   6.8 (1\%) &     59 (22\%) &   6.7 (0\%) &     58 (49\%) \\
revPBE-D3(BJ) \cite{GrimmeJCC11} &   6.5 (-4\%) &     67 (41\%) &   6.3 (-5\%) &     69 (77\%) \\
BLYP-D3(BJ) \cite{GrimmeJCC11} &   6.6 (-2\%) &     70 (46\%) &   6.5 (-3\%) &     71 (83\%) \\
\hline
MGGA+D        \\
MGGA\_MS1-D3 \cite{SunJCP13} &   6.9 (2\%) &     46 (-4\%) &   6.8 (2\%) &     44 (12\%) \\
MGGA\_MS2-D3 \cite{SunJCP13} &   6.8 (2\%) &     45 (-6\%) &   6.6 (-1\%) &     46 (17\%) \\
MGGA\_MS0-D3 \cite{SunJCP13} &   7.0 (5\%) &     42 (-13\%) &   6.8 (3\%) &     42 (7\%) \\
TPSS-D3 \cite{GrimmeJCP10} &   6.7 (0\%) &     47 (-1\%) &   6.5 (-2\%) &     50 (28\%) \\
TPSS-D3(BJ) \cite{GrimmeJCC11} &   6.5 (-3\%) &     58 (21\%) &   6.3 (-5\%) &     60 (53\%) \\
\hline
hybrid-GGA+D  \\
YSPBE0-D3(BJ) \cite{DFTD3} (0.25) &   7.0 (4\%) &     46 (-5\%) &   6.8 (2\%) &     46 (19\%) \\
PBE0-D3 \cite{GrimmeJCP10} (0.25) &   6.9 (2\%) &     41 (-14\%) &   6.7 (0\%) &     45 (16\%) \\
PBE0-D3(BJ) \cite{GrimmeJCC11} (0.25) &   6.7 (0\%) &     50 (3\%) &   6.5 (-2\%) &     51 (31\%) \\
B3LYP-D3 \cite{GrimmeJCP10} (0.20) &   6.8 (2\%) &     50 (5\%) &   6.7 (1\%) &     54 (38\%) \\
YSPBE0-D3 \cite{DFTD3} (0.25) &   7.3 (9\%) &     30 (-36\%) &   7.1 (6\%) &     30 (-23\%) \\
B3LYP-D3(BJ) \cite{GrimmeJCC11} (0.20) &   6.7 (-1\%) &     62 (29\%) &   6.5 (-3\%) &     64 (64\%) \\
\hline
hybrid-MGGA+D \\
MGGA\_MS2h-D3 \cite{SunJCP13} (0.09) &   6.8 (2\%) &     45 (-7\%) &   6.6 (-1\%) &     46 (18\%) \\
TPSSh-D3 \cite{GoerigkPCCP11} (0.10) &   6.7 (0\%) &     47 (-2\%) &   6.5 (-2\%) &     51 (30\%) \\
TPSS0-D3 \cite{GrimmeJCP10} (0.25) &   6.6 (-1\%) &     48 (0\%) &   6.5 (-3\%) &     53 (35\%) \\
TPSS0-D3(BJ) \cite{GrimmeJCC11} (0.25) &   6.5 (-3\%) &     56 (17\%) &   6.2 (-6\%) &     60 (54\%) \\
TPSSh-D3(BJ) \cite{HoffmannJCC14} (0.10) &   6.5 (-3\%) &     61 (28\%) &   6.3 (-5\%) &     63 (62\%) \\
\hline
Previous works \\
optB88-vdW \cite{KlimesJPCM10} (Ref.~\onlinecite{BjorkmanJCP14}) &  6.76 (1\%) &     66 (38\%) &  6.64 (1\%) &     67 (72\%) \\
C09x-vdW \cite{CooperPRB10} (Ref.~\onlinecite{BjorkmanJCP14}) &  6.54 (-2\%) &     71 (48\%) &  6.42 (-3\%) &     73 (87\%) \\
VV10 \cite{VydrovJCP10,SabatiniPRB13} (Ref.~\onlinecite{BjorkmanPRB12}) &  6.68 (0\%) &     71 (48\%) &  6.57 (0\%) &     70 (79\%) \\
rev-vdW-DF2 \cite{HamadaPRB14} (Ref.~\onlinecite{HamadaPRB14}) &  6.64 (-1\%) &     60 (25\%) &  6.56 (-1\%) &     57 (46\%) \\
PW86R-VV10sol \cite{BjorkmanPRB12} (Ref.~\onlinecite{BjorkmanPRB12}) &  6.98 (5\%) &     44 (-8\%) &  6.87 (4\%) &     43 (10\%) \\
AM05-VV10sol \cite{BjorkmanPRB12} (Ref.~\onlinecite{BjorkmanPRB12}) &  6.99 (5\%) &     45 (-6\%) &  6.84 (4\%) &     41 (5\%) \\
PBE+TS \cite{TkatchenkoPRL09} (Ref.~\onlinecite{BuckoPRB13}) &  6.68 (0\%) &     82 (71\%) &  6.64 (1\%) &     87 (123\%) \\
PBE+TS+SCS \cite{TkatchenkoPRL12} (Ref.~\onlinecite{BuckoPRB13}) &  6.75 (1\%) &     55 (15\%) &  6.67 (1\%) &     73 (87\%) \\
RPA (Refs. \onlinecite{LebeguePRL10,BjorkmanPRL12}) &  6.68 &     48 &  6.60 &     39 \\
\end{tabular}
\end{ruledtabular}
}
\end{table*}
\endgroup

In this section, the results for rare-gas solids (Ne, Ar, and Kr) and
layered solids (graphite and h-BN) are discussed.
Rare-gas dimers and solids, which are bound by the dispersion interactions, have been
commonly used for the testing of theoretical methods (see Refs.
\onlinecite{KullieCP12,RoyJCP12,TranJCP13,MoellmannJPCC14,CallsenPRB15,PatraTCC15} for the
most recent works). The same is true for graphite and h-BN which are stacks
of weakly bound hexagonal layers.
\cite{BjorkmanPRL12,BjorkmanJPCM12,BjorkmanPRB12,BjorkmanJCP14,GrazianoJPCM12,HamadaPRB14,BuckoPRB13,RegoJPCM15}

As mentioned in Sec. \ref{details}, the results for such weakly bound systems
are more sensitive to self-consistency effects than for the strongly bound solids.
Figure \ref{fig_Ar} shows the example for Ar, where the MGGA\_MS2
total-energy curves were obtained with four different sets of orbitals/density.
Actually, this is a particularly bad case where the
spread in the values for $a_{0}$ (5.4-5.7~\AA)
is two orders of magnitude larger than for most strongly bound solids and
the spread for $E_{\text{coh}}$ is of the same magnitude as $E_{\text{coh}}$ itself.
We observed that in general the spread in $a_{0}$ is larger for functionals
which lead to shallow minimum.
This shows that there is some non-negligible degree of uncertainty in the
results for the MGGA and hybrid functionals of Tables~\ref{results_RG} and
\ref{results_LC} that were
obtained non-self-consistently with PBE orbitals/density instead of
self-consistently as it should be.
Thus, for these functionals the
discussion should be kept at a qualitative level.
On the other hand, the main conclusions of this section should
not be affect too significantly, since for such weakly bound systems
the errors with DFT functionals are often extremely large such that only the
trends are usually discussed.

\subsubsection{\label{raregas}Rare-gas solids}

The results for the lattice constant and cohesive energy
of the rare-gas solids are shown in
Table~\ref{results_RG}. The error (indicated in parenthesis) is with respect
to accurate results obtained from the CCSD(T) method.\cite{RosciszewskiPRB00}
Also shown are results taken from Refs.
\onlinecite{AlSaidiJCTC12,TranJCP13,CallsenPRB15,HarlPRB08} that were obtained with
nonlocal dispersion-corrected functionals [Eq.~(\ref{Ecdisp2})],
the atom-pairwise method of Tkatchenko and Scheffler,\cite{TkatchenkoPRL09}
and post-PBE RPA calculations.
In a few cases, no minimum in the total-energy curve was obtained in the
range of lattice constants that we have considered (largest values are
5.6, 6.6, 7.1~\AA~for Ne, Ar, Kr). This concerns the functionals
revPBE, AM05, SG4, revTPSS(h), and those using
B88 exchange (BLYP, B3LYP, and B3PW91).
No minimum at all should exist with the B88-based functionals
(see, e.g., Refs. \onlinecite{KristyanCPL94,PerezJordaCPL95,WuJCP02,XuJCP05}), while only a very weak
minimum at a larger lattice constant could eventually be expected with
revPBE (see Ref.~\onlinecite{TkatchenkoPRB08}) and
revTPSS (see Ref.~\onlinecite{SunPRL13}).
Note that no or a very weak binding is typically obtained by
GGA functionals which violate the local Lieb-Oxford bound\cite{LiebIJQC81}
because of an enhancement factor that is too large at large
$s$ ($s\gtrsim5$) like B88 and AM05 (see Fig.~\ref{fig_Fxc1}).
The importance of the behavior of the enhancement factor at large $s$
for noncovalent interactions was underlined in
Refs. \onlinecite{WesolowskiJPCA97,ZhangJCP97}.

Unsurprisingly, the best functionals are those which include the atom-pairwise
dispersion term D3/D3(BJ), since for many of them the errors are below
$\sim8$\% for $a_{0}$ and below $\sim20$\% for $E_{\text{coh}}$ for all three
rare gases.
Such results are expected since the atom-dependent parameters in
Eq.~(\ref{Ecdisp1}) (computed almost from first principles\cite{GrimmeJCP10})
should remain always accurate in the case of interaction between rare-gas
atoms, whether it is in the dimer or in the solid. Note, however, that the
error for the cohesive energy of Ne is above 100\% for all MGGA\_MS$n$(h)-D3
functionals, which may be due to the fact that only the term $n=6$
in Eq.~(\ref{Ecdisp1}) has been considered for these functionals.\cite{SunJCP13}
All other functionals, without exception, lead to errors for $E_{\text{coh}}$
which are above 50\% for at least two rare gases. These large errors are always
due to an underestimation for Ar and Kr, but not for Ne (overestimation with
the MGGAs and underestimation with the others). For MGGA\_MS2 and SCAN, the largest
errors are $-66$\% (Ar) and 107\% (Ne), respectively.
Note that the GGA PBEfe and MGGA mBEEF overestimate the cohesive energy
even more than LDA does. For $a_{0}$, the values
obtained with the GGA PBEalpha and all modern (hybrid-)MGGAs like MGGA\_MS2, SCAN,
and mBEEF are in fair agreement with the CCSD(T) results since the errors are of
the same order as with most dispersion-corrected functionals (below 8\%). 

Concerning previous works reporting tests on other functionals, we mention
Ref.~\onlinecite{TranJCP13} where several variants of the nonlocal van der
Waals functionals were tested on rare-gas dimers (from He$_{2}$ to Kr$_{2}$)
and solids (Ne, Ar, and Kr). The conclusion was that
rVV10\cite{VydrovJCP10,SabatiniPRB13} leads to excellent results for the dimers
and is among the good ones for the solids along with optB88-vdW\cite{KlimesJPCM10}
and C09x-vdW.\cite{CooperPRB10} However, with these three nonlocal functionals
rather large errors in $E_{\text{coh}}$ were still observed for the solids
(results shown in Table~\ref{results_RG}), such that overall these functionals
are less accurate than DFT-D3/D3(BJ) for the rare-gas solids.
Another nonlocal functional, rev-vdW-DF2, was recently proposed by
Hamada,\cite{HamadaPRB14} and the results on rare-gas solids\cite{CallsenPRB15}
(see Table~\ref{results_RG}) are as good as the DFT-D3/D3(BJ) results and, therefore,
better than for the other three nonlocal functionals.
For the Ar and Kr dimers, the SCAN+rVV10 functional was shown to be as accurate as
rVV10,\cite{Peng15} however it has not been tested on rare-gas solids.
Finally, we also mention the
RPA (fifth rung of Jacob's ladder) results from Ref.~\onlinecite{HarlPRB08}
which are rather accurate overall, as shown in Table~\ref{results_RG}.
Concerning other semilocal approximations not augmented with a dispersion
correction, previous works reported unsuccessful attempts to find such a
functional leading to accurate results for all rare-gas dimers at the same time
(see, e.g., Refs.~\onlinecite{XuJCP05,ZhaoJPCA06}).

The summary for the rare-gas solids is the following. For the cohesive energy,
the functionals which include an atom-pairwise term D3/D3(BJ) clearly outperform
the others. It was also observed that the MGGAs do not improve over the GGAs
for $E_{\text{coh}}$. However, for the lattice constant, MGGAs are
superior to the GGAs and perform as well as the dispersion
corrected-functionals. Among the previous works also considering rare-gas solids
in their test set, we noted that the rev-vdW-DF2 nonlocal functional
\cite{HamadaPRB14,CallsenPRB15} shows similar accuracy as the DFT-D3/D3(BJ) methods.

\subsubsection{\label{layered}Layered solids}

Turning now to the layered solids graphite and h-BN, the results for the
equilibrium lattice constant $c_{0}$ (the interlayer distance is $c_{0}/2$)
and interlayer binding energy $E_{\text{b}}$ are shown in Table~\ref{results_LC}.
Since for these two systems we are interested only in the interlayer
properties, the intralayer lattice constant $a$ was kept fixed at the
experimental value of 2.462 and 2.503~\AA~for graphite and h-BN, respectively.
As in the recent works of Bj\"{o}rkman \textit{et al}.,
\cite{BjorkmanPRL12,BjorkmanJPCM12,BjorkmanPRB12,BjorkmanJCP14} the results from
the RPA method,\cite{LebeguePRL10,BjorkmanPRL12} which are
in very good agreement with experiment and Monte-Carlo
simulation\cite{SpanuPRL09} for graphite, are used as reference.
No experimental result for $E_{\text{b}}$ for h-BN seems to be available.

The results with the GGA functionals are extremely inaccurate since for all
these methods, except PBEfe, there is no or a very tiny binding between the
layers (underestimation of $E_{\text{b}}$ by more than 90\%) and a huge
overestimation of $c_{0}$ by at least 0.5~\AA. The underestimation of
$E_{\text{b}}$ with PBEfe is $\sim50\%$ and $c_{0}$ is too large by $\sim0.3$~\AA.
LDA also underestimates $E_{b}$ by $\sim40\%$, but leads to interlayer
distances which agree quite well with RPA and, actually, perfectly for graphite.
The best MGGA functional is MVS, whose relative error is $-34\%$ for the binding
energy of graphite, but below 5\% otherwise. SCAN performs slightly worse
since $E_{\text{b}}$ is too small by $\sim50\%$ for both graphite and h-BN,
while MGGA\_MS2 leads to disappointingly small binding energies.
The other MGGAs, including mBEEF, lead to very small (large) values for
$E_{\text{b}}$ ($c_{0}$). Note that the results obtained with mBEEF show
totally different trends as those for the rare gases (large
overbinding for the rare gases and large underbinding for graphite and h-BN),
which is maybe due to the nonsmooth form of this functional
(see Fig.~\ref{fig_Fxc2}) such that the results are more unpredictable.
Among the hybrid functionals, MVSh is the only one
which leads to somehow reasonable results, with errors for $E_{\text{b}}$ that
are slightly larger than for the underlying semilocal MVS.
Let us remark that all functionals without dispersion correction
underestimate the interlayer binding energy of graphite and h-BN, while
it was not the case for the cohesive energy of the rare-gas solids.

Adding the D3 or D3(BJ) dispersion term usually improves the agreement with
RPA, such that for many of these methods the magnitude of the relative error is
below 20\% and 3\% for $E_{\text{b}}$ and $c_{0}$, respectively. Such errors
can be considered as relatively modest. Among the computationally cheap GGA+D,
the accurate functionals are PBE-D3/D3(BJ), RPBE-D3, and PBEsol-D3.

The results obtained with many other methods are available in the literature
\cite{BjorkmanPRL12,BjorkmanJPCM12,BjorkmanPRB12,BjorkmanJCP14,GrazianoJPCM12,HamadaPRB14,BuckoPRB13,RegoJPCM15}
(see Ref.~\onlinecite{RegoJPCM15} for a collection of values for graphite),
and those obtained from the methods that were already selected for the discussion on
the rare gases are shown in Table~\ref{results_LC}. The nonlocal functionals
optB88-vdW, C09x-vdW, VV10, and rev-vdW-DF2 as well as PBE+TS(+SCS) lead to
very good agreement with RPA for the interlayer lattice constant (errors in the
range 0-2\%), however, in most cases there is a non-negligible
overbinding above 40\%. Also shown in Table~\ref{results_LC}, are the results
obtained with the nonlocal functionals PW86R-VV10 and AM05-VV10sol which contain
parameters that were fitted specifically to RPA binding energies of 26 layered solids
including graphite and h-BN.\cite{BjorkmanPRB12} Unsurprisingly, the errors obtained with
PW86R-VV10 and AM05-VV10sol for $E_{\text{b}}$ are very small (below 10\%),
but the price to pay are errors for $c_{0}$ that are clearly larger ($\sim5\%$)
than with the other nonlocal
functionals. The nonlocal functional SCAN+rVV10 has also been tested on a
set of 28 layered solids, and according to Ref.~\onlinecite{Peng15}, the
MARE (the detailed results for each system are not available) for the
interlayer lattice constant and binding energy amount to
1.5\% and 7.7\%, respectively, meaning that SCAN+rVV10 leads to very low
errors for both quantities, despite no parameter was tuned
to reproduce the results for the layered solids.

In summary, among the methods which do not include an atom-pairwise dispersion
correction, only a couple of them (MVS, LDA, and PBEfe) do not severely
underestimate the interlayer binding energy. Adding a D3/D3(BJ) atom-pairwise
term clearly improves the results, leading to rather satisfying values for the
interlayer spacing and binding energy. In the group of nonlocal functionals, the
recently proposed SCAN+rVV10 seems to be among the most accurate.\cite{Peng15}

\section{\label{literature}Brief overview of literature results for molecules}

The results that have been presented and discussed so far
concern exclusively solid-state properties and may certainly not reflect the
trends for finite systems, as mentioned in Sec.~\ref{functionals}. Thus, in
order to provide to the reader of the present work a more general view on the accuracy and
applicability of the functionals, a very brief summary of some of the
literature results for molecular systems is given below. To this end we
consider the atomization energy of strongly bound molecules and the interaction
energy between weakly bound molecules, for which widely used
standard testing sets exist.

\subsection{\label{molecules}Atomization energy of molecules}

The atomization energy of molecules is one of the most used quantity to assess
the performance of functionals for finite systems (see, e.g., Refs.
\onlinecite{PeveratiPCCP12a,WellendorffPRB12,SunJCP13,SunPRL15,MardirossianJCP15,FabianoTCA15}
for recent tests). Large testing sets of small molecules like G3\cite{CurtissJCP00}
or W4-11\cite{KartonCPL11} usually involve only elements of the first
three periods of the periodic table. For such sets, the MAE given by LDA is
typically in the range 70-100 kcal/mol (atomization energies are usually
expressed in these units), while the best GGAs (e.g., BLYP)
can achieve a MAE in the range 5-10 kcal/mol. MGGAs and hybrid can reduce
further the MAE below 5 kcal/mol. At the moment, the most accurate
functionals lead to MAE in the range 2.5-3.5 kcal/mol, e.g.,
mBEEF\cite{WellendorffJCP14} (see also
Refs. \onlinecite{PeveratiPCCP12a,MardirossianJCP15}), which is not far
from the so-called \textit{chemical accuracy} of 1~kcal/mol.
Concerning the best functionals for the
solid-state test sets that we have identified just above, the MGGAs MGGA\_MS2
and SCAN, as well as the hybrid-MGGAs MGGA\_MS2h and revTPSSh are also excellent
for the atomization energy of molecule since they lead to rather small MAE around
5 kcal/mol.\cite{CsonkaJCTC10,SunJCP13,SunPRL15} With the hybrid YSPBEsol0
($\sim$HSEsol), which was the best functional for the lattice constant,
the MAE is larger (around 10-15 kcal/mol\cite{SchimkaJCP11}).
The weak GGAs WC and PBEsol improve only slightly over LDA since their MAE are
as large as 40-60 kcal/mol,\cite{WellendorffPRB12,SunPRL15} while
the stronger GGA PBEint leads to a MAE around
20-30 kcal/mol.\cite{FabianoIJQC13}

Therefore, as mentioned in Sec.~\ref{functionals}, it really seems that the
kinetic-energy density is a necessary ingredient in order to construct a
functional that is among the best for both solid-state properties and the
atomization energy of molecules, and some of the modern MGGAs like SCAN look
promising in this respect. With GGAs, it looks like an unachievable task to get
such universally good results. Hybrid-GGAs can improve upon the underlying GGA,
however we have not been able to find an excellent functional in our test set.
For instance, YSPBEsol ($\sim$HSEsol) is very good for solids, but not for
molecules, while the reverse is true for PBE0 (small MAE of $\sim7$ kcal/mol
for molecules,\cite{StaroverovJCP03} but average results for solids, see
Table~\ref{results_strong}).

\subsection{\label{S22}S22 set of noncovalent complexes}

\begin{table}
\caption{\label{results_S22}Results from the literature
(reference in last column) for the MAE
(in kcal/mol) on the S22 testing set.}
\begin{ruledtabular}
\begin{tabular}{lcc}
Functional & MAE & Reference \\
\hline
LDA & & \\
LDA\cite{PerdewPRB92a} & 2.3 & \onlinecite{SunPRL15} \\
\hline
GGA & & \\
PBEsol\cite{PerdewPRL08} & 1.8 & \onlinecite{SunPRL15} \\
PBE\cite{PerdewPRL96} & 2.8 & \onlinecite{SunPRL15} \\
RPBE\cite{HammerPRB99} & 5.2 & \onlinecite{WellendorffJCP14} \\
revPBE\cite{ZhangPRL98} & 5.3 & \onlinecite{GrimmeJCP10} \\
BLYP\cite{BeckePRA88,LeePRB88} & 4.8, 8.8 & \onlinecite{GrimmeJCC11}, \onlinecite{SunPRL15} \\
\hline
MGGA & & \\
M06-L\cite{ZhaoJCP06} & 0.7 & \onlinecite{MaromJCTC11} \\
MVS\cite{SunPNAS15} & 0.8 & \onlinecite{SunPNAS15} \\
SCAN\cite{SunPRL15} & 0.9 & \onlinecite{SunPRL15} \\
mBEEF\cite{WellendorffJCP14} & 1.4 & \onlinecite{WellendorffJCP14} \\
MGGA\_MS0\cite{SunJCP12} & 1.8 &\onlinecite{WellendorffJCP14} \\
MGGA\_MS2\cite{SunJCP13} & 2.1 & \onlinecite{WellendorffJCP14} \\
revTPSS\cite{PerdewPRL09} & 3.4 & \onlinecite{WellendorffJCP14} \\
TPSS\cite{TaoPRL03} & 3.7 & \onlinecite{SunPRL15} \\
\hline
hybrid-GGA & & \\
HSE\cite{HeydJCP03} & 2.4 & \onlinecite{MaromJCTC11} \\
PBE0\cite{ErnzerhofJCP99,AdamoJCP99} & 2.5 & \onlinecite{MaromJCTC11} \\
B3LYP\cite{StephensJPC94} & 3.8 & \onlinecite{GrimmeJCP10} \\
\hline
hybrid-MGGA & & \\
MVSh\cite{SunPNAS15} & 1.0 & \onlinecite{SunPNAS15} \\
M06\cite{ZhaoTCA08} & 1.4 & \onlinecite{MaromJCTC11} \\
TPSS0\cite{TaoPRL03,PerdewJCP96} & 3.1 & \onlinecite{GrimmeJCP10} \\
\hline
GGA+D & & \\
BLYP-D3\cite{GrimmeJCP10} & 0.2 & \onlinecite{GrimmeJCC11} \\
BLYP-D3(BJ)\cite{GrimmeJCC11} & 0.2 & \onlinecite{GrimmeJCC11} \\
PBE+TS\cite{TkatchenkoPRL09}\footnotemark[1] & 0.3 & \onlinecite{MaromJCTC11} \\
PW86PBE-XDM(BR)\cite{KannemannJCTC10} & 0.3 & \onlinecite{KannemannJCTC10} \\
revPBE-D3\cite{GrimmeJCP10} & 0.4 &\onlinecite{GrimmeJCC11} \\
revPBE-D3(BJ)\cite{GrimmeJCC11} & 0.4 & \onlinecite{GrimmeJCC11} \\
PBE-D3\cite{GrimmeJCP10} & 0.5 & \onlinecite{GrimmeJCC11} \\
PBE-D3(BJ)\cite{GrimmeJCC11} & 0.5 & \onlinecite{GrimmeJCC11} \\
\hline
MGGA+D & & \\
TPSS-D3\cite{GrimmeJCP10} & 0.3 & \onlinecite{GrimmeJCC11} \\
TPSS-D3(BJ)\cite{GrimmeJCC11} & 0.3 &\onlinecite{GrimmeJCC11} \\
MGGA\_MS0-D3\cite{SunJCP13}\footnotemark[1] & 0.3 & \onlinecite{SunJCP13} \\
MGGA\_MS1-D3\cite{SunJCP13}\footnotemark[1] & 0.3 & \onlinecite{SunJCP13} \\
MGGA\_MS2-D3\cite{SunJCP13}\footnotemark[1] & 0.3 & \onlinecite{SunJCP13} \\
\hline
hybrid-GGA+D & & \\
B3LYP-D3(BJ)\cite{GrimmeJCC11} & 0.3 & \onlinecite{GrimmeJCC11} \\
B3LYP-D3\cite{GrimmeJCP10} & 0.4 & \onlinecite{GrimmeJCC11} \\
PBE0-D3(BJ)\cite{GrimmeJCC11} & 0.5 & \onlinecite{GrimmeJCC11} \\
PBE0-D3\cite{GrimmeJCP10} & 0.6 & \onlinecite{GrimmeJCC11} \\
\hline
hybrid-MGGA+D & & \\
MGGA\_MS2-D3\cite{SunJCP13}\footnotemark[1] & 0.2 & \onlinecite{SunJCP13} \\
TPSS0-D3\cite{GrimmeJCP10} & 0.4 & \onlinecite{GrimmeJCC11} \\
TPSS0-D3(BJ)\cite{GrimmeJCC11} & 0.4 & \onlinecite{GrimmeJCC11} \\
\hline
GGA+NL & & \\
optB88-vdW\cite{KlimesJPCM10}\footnotemark[1] & 0.2 &\onlinecite{KlimesJPCM10} \\
C09x-vdW\cite{CooperPRB10} & 0.3 & \onlinecite{CooperPRB10} \\
VV10\cite{VydrovJCP10}\footnotemark[1] & 0.3 & \onlinecite{VydrovJCP10} \\
rev-vdW-DF2\cite{HamadaPRB14} & 0.5 & \onlinecite{HamadaPRB14} \\
vdW-DF\cite{DionPRL04} & 1.5 & \onlinecite{KlimesJPCM10} \\
\hline
MGGA+NL & & \\
SCAN+rVV10\cite{Peng15} & 0.4 & \onlinecite{Peng15} \\ 
BEEF-vdW\cite{WellendorffPRB12} & 1.7 & \onlinecite{WellendorffJCP14} \\
\end{tabular}
\end{ruledtabular}
\footnotetext[1]{One or several parameters were determined using the S22 set.}
\end{table}

The S22 set of molecular complexes,\cite{JureckaPCCP06} which consists of 22
dimers of biological-relevance molecules bound by weak interactions
(hydrogen-bonded, dispersion dominated, and mixed), has become a standard set
for the testing of functionals since very accurate CCSD(T) interaction energies
are available.\cite{JureckaPCCP06,TakataniJCP10,PodeszwaPCCP10}
A large number of functionals have already been assessed on the S22 set, and
Table~\ref{results_S22} summarizes the results taken from the literature for
many of the functionals that we have considered in the present work.
Also included are results for nonlocal van der Waals functionals
(groups GGA+NL and MGGA+NL), two atom-pairwise dispersion methods
[PBE+TS\cite{TkatchenkoPRL09} and PW86PBE-XDM(BR)\cite{KannemannJCTC10}],
and the highly parameterized Minnesota functionals M06 and M06-L\cite{ZhaoTCA08}
(results for other highly parameterized functionals can be found in
Refs. \onlinecite{MardirossianPCCP14,MardirossianJCP14,MardirossianJCP15}).
Since all these recent results are widely scattered in the literature,
it is also timely to gather them in a single table (see also
Ref.~\onlinecite{GrimmeWCMS11}). As indicated in Table~\ref{results_S22}, some
of the functionals contain one or several parameters that were fitted using the
CCSD(T) interaction energies of the S22 set.

From the results, it is rather clear that
the dispersion-corrected functionals are more accurate. The MAE is usually in
the range 0.2-0.5 kcal/mol, while it is above 1~kcal/mol for the methods
without dispersion correction term, except M06-L, MVS and SCAN (slightly below
1~kcal/mol). As already observed for the layered compounds in
Sec.~\ref{layered}, MVS is one of the best non-dispersion corrected
functionals, which is probably due to the particular form of the enhancement
factor that is a strongly decreasing function of $s$ and $\alpha$ (see
Fig.~\ref{fig_Fxc2}). The same can be said about SCAN, which is also one
of the semilocal functionals which do not completely and systematically fail
for weak interactions. The MAE obtained with MGGA\_MS2 is rather large
(2.1~kcal/mol), despite it was the best MGGA for the rare-gas solids.
Among the GGAs, PBEsol represents a good balance between LDA and PBE which
overestimate and underestimate the interaction energies,
respectively,\cite{SunPRL15} but leads to a MAE which is still rather high
(1.8~kcal/mol). In the group of nonlocal vdW functionals, the early vdW-DF and
recent BEEF-vdW are clearly less accurate (MAE around 1.5~kcal/mol) than the
others like SCAN+rVV10. The largest MAE obtained with an atom-pairwise
dispersion method is only 0.6~kcal/mol (PBE0-D3).

By considering all results for weak interactions discussed in this work
(rare-gas solids, layered solids, and S22 molecules), the most important
comments are the following. For the three sets of systems, the
atom-pairwise methods show a clear improvement over the other methods.
Such an improvement is (slightly) less visible with the nonlocal
vdW functionals, especially for the rare-gas and layered solids,
where only rev-vdW-DF2 seems to compete with the best atom-pairwise methods.
In Ref.~\onlinecite{Peng15}, the recent SCAN+rVV10 nonlocal functional has
shown to be very good for layered solids and the molecules of the S22 test set,
but no results for rare-gas solids are available yet.
Among the non-dispersion corrected functionals, only a few lead
\textit{occasionally} to more or less reasonable results. This concerns mainly
the recent MGGA functionals MGGA\_MS2, MVS, and SCAN, however, their accuracies
are still clearly lower than the atom-pairwise methods.

\section{\label{summary}Summary}

A large number of exchange-correlation functionals have been tested for
solid-state properties, namely, the lattice constant, bulk modulus, and cohesive
energy. Functionals from the first four rungs of Jacob's ladder were considered
(i.e., LDA, GGA, MGGA, and hybrid) and some of them were augmented
with a D3/D3(BJ) term to account explicitly for the dispersion interactions.
The testing set of solids was divided into two groups: the solids bound by
strong interactions (i.e., covalent, ionic, or metallic) and those bound
by weak interactions (e.g., dispersion). Furthermore, in order to give a broader
view of the performance of some of the tested functionals, a section was
devoted to a summary of the literature results on molecular systems for
two properties; the atomization energy and intermolecular binding.

One of the purposes of this work was to assess the accuracy of some of the
recently proposed functionals like the MGGAs MGGA\_MS$n$, SCAN,
and mBEEF, and to identify, eventually, an \textit{universally good} functional.
Another goal was to figure out how useful it is to mix HF exchange with
semilocal exchange or to add a dispersion correction term. An attempt to
provide an useful summary of the most important observations of this work is
the following:
\begin{enumerate}
\item For the strongly bound solids (Table~\ref{results_strong}), at least
one functional of each rung of Jacob's ladder, except LDA, belongs to the
group of the most accurate functionals. Although it is not always obvious
to decide if a functional should be a member of this group or not, we can mention
the GGAs WC, PBEsol, PBEalpha, PBEint, and SG4, the MGGAs MGGA\_MS2 and SCAN,
the hybrids YSPBEsol0, MGGA\_MS2h, and revTPSSh, and the dispersion corrected
methods PBE-D3/D3(BJ).
\item Thus, from point 1 it does not seem to be really necessary to
go beyond the GGA approximation for strongly bound solids since a few
of them are overall as accurate as the more sophisticated/expensive
MGGA and hybrid functionals. However, the use of a MGGA or hybrid
functional may be necessary for several reasons as explained in points 3 and
4 below.
\item As well known (see Sec.~\ref{molecules}), no GGA can be excellent
for both solids and molecules at the same time. MGGAs like MGGA\_MS2 or SCAN
are better in this respect. Thus, the use of a MGGA should be more recommended
for systems involving both finite and infinite systems as exemplified
in, e.g., Ref.~\onlinecite{SunPRB11a}.
\item If a qualitatively more accurate prediction of the band gap of
semiconductors and insulators is also required, then an hybrid functional
should be used since GGA band gaps are usually by far too small compared
to experiment.\cite{HeydJCP05,SchimkaJCP11,PernotJPCA15}
Note, however, that hybrid functionals are not recommended
for metallic systems.\cite{TranPRB12,JangJPSJ12,Gao15}
MGGAs do not really improve over GGAs for the band gap
(see Refs. \onlinecite{ZhaoJCP09,XiaoPRB13,Yang16}).
\item The use of a dispersion correction for strongly bound solids
is recommended for functionals which clearly overestimate the lattice
constant (usually more pronounced for solids containing alkali atoms).
However, it is only in the case of PBE-D3/D3(BJ) that the overall accuracy
is really good. We also observed that many of the DFT-D3/D3(BJ) methods lead to
large errors for the bulk modulus, despite small errors for the lattice constant.
\item Among the functionals that were not tested in the present work,
SCAN+rVV10 should be of similar accuracy as SCAN for strongly bound solids,
but significantly improves the results for weak interactions according to
Ref.~\onlinecite{Peng15}.
\item For the weakly bound systems, namely the rare-gas solids
(Table~\ref{results_RG}), layered solids (Table~\ref{results_LC}), and
intermolecular complexes (Table~\ref{results_S22}), it was observed
that none of the non-dispersion corrected functionals is able to
give qualitatively correct results in most cases. This is expected
since the physics of dispersion is not included in the construction of
these functionals. At best, good results can occasionally be expected
with some of the MGGAs (MGGA\_MS2, MVS, SCAN, or M06-L).
\item For weak interactions, many of the DFT-D3/D3(BJ) methods (and some
of the nonlocal ones) are much more reliable. The results from the
literature obtained with the recent SCAN+rVV10\cite{Peng15}
or B97M-V\cite{MardirossianJCP15} are promising, but since these functionals were
proposed very recently, more tests are needed in order to have a more
complete view of their general performance.
\end{enumerate}
Finally, from the present and previously published works, a very short
conclusion would be the following. At the present time, it seems that the only
functionals which can be among the most accurate for the geometry and
energetics in \textit{both} finite and infinite systems and for \textit{both}
strong and weak bondings, are MGGAs augmented with a dispersion term.
These are not bad news, since MGGAs functionals are barely more expensive than
GGAs and the addition of a pairwise or nonlocal dispersion term does not
significantly increases the computational time.
If qualitatively accurate band gaps are also needed, then such functionals
should be mixed with HF exchange, but with the disadvantage of a
significantly increased computational cost, especially for large molecules
or periodic solids.

\begin{acknowledgements}

This work was supported by the project SFB-F41 (ViCoM) of the Austrian Science Fund.

\end{acknowledgements}

\bibliography{/area52/tran/divers/references}

\end{document}